\documentclass[twocolumn,showpacs,showkeys,superscriptaddress]{revtex4}
\usepackage{amsmath}
\usepackage{amssymb}
\usepackage{graphicx}

\newcommand{\bastar}{\begin{eqnarray*}}
\newcommand{\eastar}{\end{eqnarray*}}
\newskip\humongous \humongous=0pt plus 1000pt minus 1000pt

\relax
\newcommand{\bea}{\begin{eqnarray}}
\newcommand{\eea}{\end{eqnarray}}
\newcommand{\X}{{\vec X}}
\newcommand{\pro}{\partial}
\newcommand{\hn}{\hat n}
\newcommand{\hr}{\hat r}
\newcommand{\pd}{\partial}
\newcommand{\mn}{{\mu \nu}}
\newcommand{\hD}{{\hat D}}
\newcommand{\n}{\vec n}

\newcommand{\A}{{\vec A}}

\newcommand{\ba}{\begin{array}}
\newcommand{\ea}{\end{array}}
\newcommand{\Int}{{\displaystyle \int}}

\newcommand{\nn}{\nonumber}
\newcommand{\om}{\omega}
\begin{document}
\title{Re-interpretation of Skyrme Theory: New Topological
structures}
\author{Pengming Zhang}
\affiliation{Institute of Modern Physics, Chinese Academy
of Science, Lanzhou 730000, China}
\author{Kyoungtae Kimm}
\affiliation{Faculty of Liberal Education,
Seoul National University, Seoul 151-747, Korea}
\author{Liping Zou}
\affiliation{Institute of Modern Physics, Chinese Academy
of Science, Lanzhou 730000, China}
\affiliation{Institute of Basic Science,
Konkuk University, Seoul 143-701, Korea}
\author{Y. M. Cho}
\email{ymcho7@konkuk.ac.kr}
\affiliation{Institute of Modern Physics, Chinese Academy
of Science, Lanzhou 730000, China}
\affiliation{Administration Building 310-4,
Konkuk University, Seoul 143-701, Korea}
\affiliation{School of Physics and Astronomy,
Seoul National University, Seoul 151-742, Korea}

\begin{abstract}
~~~~~Recently it has been pointed out that the skyrmions 
carry two independent topology, the baryon topology and 
the monopole topology. We provide more evidence to support 
this. In specific, we prove that the baryon number $B$ can 
be decomposed to the monopole number $m$ and the shell 
number $n$, so that $B$ is given by $B=mn$. This tells that 
the skyrmions may more conveniently be classified by two 
integers $(m,n)$. This is because the rational map which 
determines the baryon number in the popular multi-skyrmion 
solutions actually describes the monopole topology 
$\pi_2(S^2)$ which is different from the baryon topology 
$\pi_3(S^3)$. Moreover, we show that the baby skyrmions 
can also be generalized to have two topology, $\pi_1(S^1)$ 
and $\pi_2(S^2)$, and thus should be classified by two 
topological numbers $(m,n)$. Furthermore, we show that 
the vacuum of the Skyrme theory can be classified by 
two topological numbers $(p,q)$, the $\pi_1(S^1)$ of 
the sigma-field and the $\pi_3(S^2)$ of the normalized 
pion-field. This means that the Skyrme theory has multiple 
vacua similar to the Sine-Gordon theory and QCD combined 
together. This puts the Skyrme theory in a totally new 
perspective. We discussthe physical implications of our 
results.
\end{abstract}
\maketitle

\section{Introduction}

The Skyrme theory has played an important role in physics.
Originally it was proposed as a theory of pion physics in
strong interaction where the skyrmion, a topological soliton
made of pions, appears as the baryon \cite{sky}. Soon after,
the theory has been interpreted as a low energy effective
theory of QCD in which the massless pion fields emerge
as the Nambu-Goldstone field of the spontaneous chiral
symmetry breaking \cite{witt,rho,bogo,prep}. This view
has become popular and very successful.

After the Skyrme's proposal novel way to obtain
multi-skyrmions without the spherical symmetry was
developed, and the skyrmions with baryon number up
to 9 based on the rational map have been constructed
and associated with real nuclei \cite{bat,man}.
Moreover, a systematic approach utilizing the shell
structure which made the numerical construction of
skyrmions with large baryon number possible has been
developed \cite{houg,sut,piet1,piet2}. This, with the improved
computational power has made people construct skyrmions
with the baryon number up to 108 \cite{sky108}.

With the new development the Skyrme theory has been able
to provide a quantitative understanding of the spectrum
of rotational excitations of carbon-12, including the Hoyle
state which is essential for the generation of heavy nuclear
elements in early universe \cite{hoyle,epel,lau}. And the
spin-orbit interaction which is essential for the magic
number of nuclei is investigated within the framework of
Skyrme theory \cite{hal}. Moreover, a method to reduce
the binding energy of skyrmions to a realistic level to
improve the Skyrme model has been developed \cite{adam}.
So by now in principle one could construct all nuclei as
multi-baryon skyrmions and discuss the phenomenology
of nuclear physics.

But the Skyrme theory has multiple faces which has
other topological objects. In addition to the well known
skyrmion it has the (helical) baby skyrmion and
the Faddeev-Niemi knot. Most importantly, it has
the monopole which plays the fundamental role. In fact
it can be viewed as a theory of monopole which has
a built-in Meissner effect \cite{prl01,plb04,ijmpa08}.
In this view all finite energy topological objects in
the theory could be viewed either as dressed monopoles
or as confined magnetic flux of the monopole-antimonopole
pair. The skyrmion can be viewed as a dressed monopole,
the baby skyrmion as a magnetic vortex created by
the monopole-antimonopole pair infinitely separated
apart, and the Faddeev-Niemi knot as a twisted magnetic
vortex ring made of the helical baby skyrmion.

This tells that the theory can be interpreted as a theory
of monopole in which the magnetic flux of the monopoles
are confined and/or screened. Indeed this has made
the theory very important not only in high energy physics
but also in condensed matter physics, in particular in
two-gap superconductor and two-component Bose-Einstein
condensates \cite{plb04,ijmpa08,ruo,pra05,manko}.

Of course, the fact that the skyrmion is closely related
to the monopole has been appreciated for a long time. It
has been well known that the rational map which played
the crucial role in the construction of the multi-skyrmions
is exactly the $\pi_2(S^2)$ mapping which provides
the monopole quantum number \cite{houg}. Nevertheless
the skyrmions have always been classified by the baryon
number given by $\pi_3(S^3)$, not by the monopole
number $\pi_2(S^2)$. This was puzzling.

This raises a serious question on the popular view
that the skyrmion can be identified as the baryon.
If the skyrmions are dressed monopoles which have
the monopole topology, one might wonder what is
the role of the monopole in the baryon.  More bluntly,
we have to explain what is the connection between
the monopole topology $\pi_2(S^2)$ and the baryon
topology $\pi_3(S^3)$ which defines the baryon
number $B$.

Besides, the Faddeev-Niemi knots add another problem.
If we adopt the popular view the Faddeev-Niemi
knots should be interpreted as topologically stable
mesons made of baryon-antibaryon pair, since they
could be viewed as the twisted vortex rings made of
skyrmion-antiskyrmion pairs. And they are expected
to have huge mass, simply because of the geometric
shape \cite{prl01,plb04,ijmpa08}. This implies
the existence of supermassive topologically stable
mesons which might be very difficult to accommodate
from the phenomenological point of view.

Recently a new proposal was made which could shed
a new light on these problems. First, it was argued
that the skyrmions could actually be classified by two
topological numbers, the baryon number $b$ and
the monopole number $m$ \cite{epjc17}. According
to this proposal the baryon number could be replaced
by the radial (shell) quantum number $n$ which
describes the $\pi_1(S^1)$ topology of the radial
extension of the skyrmions. This was based on
the observation that the $S^3$ space (both
the compactified 3-dimensional real space and
the SU(2) target space) has the Hopf fibering
$S^3\simeq S^2\times S^1$, so that the baryon
number defined by $\pi_3(S^3)$ can be decomposed
to two topological numbers $\pi_2(S^2)$ and
$\pi_1(S^1)$.

Moreover, it has been argued that the Skyrme
theory in fact has the multiple vacua similar
to the Sine-Gordon theory, each of which has
the knot topology of the QCD vacuum \cite{epjc17}.
In other words, the vacuum of the Skyrme theory
has the topology of the Sine-Gordon theory and
QCD combined together. If so, the vacuum of
the Skyrme theory can also be classified by two
integers by $(p,q)$, $p$ which represents
the $\pi_1(S^1)$ topology and $q$ which
represents the $\pi_3(S^2)$ topology.

These new features, in particular the fact that
the skyrmions can be classified by two topological
quantum numbers, naturally clarifies the relation
between the monopole topology and the baryon
topology of the skyrmion. As importantly, the fact
that the Skyrme theory has a non-trivial vacuum
topology strongly suggests that the Skyrme theory
really need a totally new interpretation.

On the other hand, these new features are totally
unexpected, and many peoples might still be skeptical
about this. The purpose of this paper is to provide
more evidence to support these new features of 
Skyrme theory, to clarify them in more detail, and 
to discuss their physical implications. In particular, 
we present new solutions of baby skyrmion, and 
show that (not only the skyrmion but also) the baby 
skyrmion has two topology, the shell (radial) topology 
$\pi_1(S^1)$ and the monopole topology $\pi_2(S^2)$.
This means that the baby skyrmion should also be 
classified by two topological numbers. This is really 
remarkable, which strongly support that the skyrmion 
does carry two topology.

The paper is organized as follows. In Section II we briefly
review the old skyrmions for later purpose. In Section III
we reveal the hidden connection between the Skyrme
theory and QCD, and emphasize that the skyrmion is
the dressed Wu-Yang monopole of QCD transplanted
in the Skyrme theory. In Section IV we show that
the skyrmions carry two topological quantum numbers,
the baryon number $b$ and the monopole number
$m$. Moreover, we show that the baryon number can
be replaced by the radial (shell) quantum number $n$,
so that they can be classified by $(m,n)$ In this scheme
the baryon number is given by $b=mn$. In Section
V we obtain new baby skyrmion solutions, and show 
how the baby skyrmion can be generalized to carry
two topology, the radial topology $\pi_1(S^1)$ and 
the monopole topology $\pi_2(S^2)$. With this we 
present the baby skyrmion solutions which carry two 
topological numbers. In SectionVI we discuss
the vacuum structure of the Skyrme theory, and
show that it has the structure of the vacuum of
the Sine-Gordon theory combined with the vacuum
of the SU(2) QCD. This tells that it can be classified
by two topological quantum numbers denoted by $(p,q)$,
where $p$ and $q$ represent the $\pi_1(S^1)$ topology
of the Sine-Gordon theory and the $\pi_3(S^2)$ topology
of QCD vacuum. Finally in Section VI we discuss the physical
implications of our results.

\section{Skyrme Theory: A Review}

Before we show that the skyrmions can be classified by
two quantum numbers, we first review the well known
facts in Skyrme theory. Let $\omega$ and $\hn$
(${\hn}^2 = 1$) be the massless scalar field
(the ``sigma-field") and the normalized pion-field
in Skyrme theory, and write the Skyrme Lagrangian
as \cite{sky}
\bea
&{\cal L} = \dfrac{\kappa^2}{4} {\rm tr} ~L_\mu^2
+ \dfrac{\alpha}{32}{\rm tr}
\left( \left[ L_\mu, L_\nu \right] \right)^2 \nn\\
&= - \dfrac{\kappa^2}{4} \Big[ \dfrac{1}{2} (\pd_\mu \om)^2
+2 \sin^2 \dfrac{\om}{2} (\pd_\mu \hn)^2 \Big]  \nn\\
&-\dfrac{\alpha}{8} \Big[ \sin^2 \dfrac{\om}{2}
\big((\pd_\mu \om)^2 (\pd_\nu \hn)^2
-(\pd_\mu \om \pd_\nu \om)
(\pd_\mu \hn \cdot \pd_\nu \hn) \big)  \nn\\
&+2 \sin^4 \dfrac{\om}{2}
(\pd_\mu \hn \times \pd_\nu \hn)^2 \Big], \nn\\
&L_\mu = U\pd_\mu U^{\dagger}, \nn\\
&U = \exp (\dfrac{\om}{2i} \vec \sigma \cdot \hn)
= \cos \dfrac{\om}{2} - i (\vec \sigma \cdot \hn)
\sin \dfrac{\om}{2},
\label{slag}
\eea
where $\kappa$ and $\alpha$ are the coupling constants.
The Lagrangian has a hidden $U(1)$ gauge symmetry
which leaves $\hn$ invariant as well as a global
$SU(2)_L \times SU(2)_R$ symmetry.

Notice that, with
\bea
&\sigma=\cos \dfrac{\omega}{2},
~~~\vec \pi= \hn \sin \dfrac{\omega}{2},
~~~(\sigma^2 + \vec \pi^2 = 1),
\eea
the Lagrangian (\ref{slag}) can also be put into the form
\bea
&{\cal L} = -\dfrac{\kappa^2}{2} \big((\partial_\mu \sigma)^2
+(\partial_\mu \vec \pi)^2 \big) \nn\\
&-\dfrac{\alpha}{4} \big((\partial_\mu \sigma \partial_\nu \vec \pi
- \partial_\nu \sigma \partial_\mu \vec \pi)^2
+ (\partial_\mu \vec \pi \times \partial_\nu \vec \pi)^2 \big) \nn\\
&+\dfrac{\lambda}{4} (\sigma^2 + \vec \pi^2 - 1),
\label{smlag}
\eea
where $\lambda$ is a Lagrange multiplier. In this form $\sigma$
and $\vec \pi$ represent the sigma field and the pion field,
so that the theory describes the non-linear sigma model of
the pion physics.

The Lagrangian has a hidden $U(1)$ gauge symmetry
as well as the global $SU(2)_L \times SU(2)_R$
symmetry \cite{plb04,ijmpa08}. The global
$SU(2)_L \times SU(2)_R$ symmetry is obvious, but
the hidden $U(1)$ gauge symmetry is not. The hidden
$U(1)$ gauge symmetry comes from the $U(1)$ subgroup
which leaves $\hn$ invariant. To see this, we
reparametrize $\hn$ by the $CP^1$ field $\xi$,
\bea
\n = \xi^\dag \vec \sigma \xi,
~~~~~\xi^\dag \xi=1.
\label{ndef}
\eea
and find that under the $U(1)$ gauge transformation
of $\xi$ to
\bea
\xi \rightarrow \exp (i\theta(x)) \xi,
\label{u1}
\eea
$\hn$ (and $\pro_\mu \hn$) remains invariant.
Now, we introduce the composite gauge potential
$C_\mu$ and the covariant derivative $D_\mu$
which transforms gauge covariantly under (\ref{u1})
by
\begin{gather}
C_\mu =-2i \xi^\dagger \pd_\mu \xi,
~~~D_\mu \xi = (\pd_\mu -\frac{i}{2} C_\mu )\xi.
\label{mpot}
\end{gather}
With this we have the following identities,
\begin{gather}
(\pd_\mu \hn )^2 = 4 |D_\mu \xi |^2,  \nn\\
\pd_\mu \hn \times \pd_\nu \hn 
=-2i \Big[(\pd_\mu \xi^\dagger)(\pd_\nu \xi) 
- (\pd_\mu \xi^\dagger)(\pd_\nu\xi)\Big] \hn \nn\\
= H_\mn \hn,  \nn\\
H_\mn=\pd_\mu C_\nu - \pd_\nu C_\mu.
\end{gather}
Furthermore, with the Fierz' identity
\begin{gather}
\sigma_{ij}^a \sigma_{kl}^a = 2\delta_{il}\delta_{jk}
- \delta_{ij}\delta_{kl},
\end{gather}
we have
\begin{gather}
\pd_\mu \hn \cdot \pd_\mu \hn
=2 \pd_\mu (\xi^\dagger_i \xi_j)
\pd_\nu (\xi^\dagger_j \xi_i)  \nn\\
= 2\Big[ (\pd_\mu\xi^\dagger\xi )(\pd_\nu \xi^\dagger\xi)
+(\pd_\mu \xi^\dagger)( \pd_\nu \xi )  
+(\partial_\nu \xi^\dagger)( \partial_\mu \xi)  \nn\\
+(\xi^\dagger \pd_\mu \xi)(\xi^\dagger \pd_\nu \xi) \Big] 
=2\Big[(\pd_\mu \xi^\dagger + \frac{i}{2} C_\mu \xi^\dagger )
(\pd_\nu \xi -\frac{i}{2} C_\nu \xi)   \nn\\
+(\pd_\nu \xi^\dagger + \frac{i}{2} C_\nu \xi^\dagger )
(\pd_\mu \xi - \frac{i}{2} C_\mu \xi) \Big] \nn\\
= 2\Big[ (D_\mu \xi )^\dagger (D_\nu \xi)
+ (D_\nu \xi )^\dagger (D_\mu \xi)\Big].
\end{gather}
From this we can express (\ref{slag}) by
\begin{gather}
{\cal L} = - \dfrac{\kappa^2}{4}
\Big[ \dfrac{1}{2} (\pd_\mu \om)^2
+8 \sin^2 \dfrac{\om}{2} |D_\mu \xi|^2 \Big]  \nn\\
-\dfrac{\alpha}{2}  \sin^2 \dfrac{\om}{2}
\Big[(\pd_\mu \om)^2 |D_\mu \xi|^2
-(\pd_\mu \om \pd_\nu \om)
(D_\mu \xi )^\dagger (D_\nu \xi) \nn\\
+\frac{1}{2}\sin^2 \dfrac{\om}{2} H_\mn^2  \Big],
\end{gather}
which is explicitly invariant under the $U(1)$ gauge
transformation (\ref{u1}). So replacing $\hn$ by
$\xi$ in the Lagrangian we can make the hidden
$U(1)$ gauge symmetry explicit. In this form
the Skyrme theory becomes a self-interacting
$U(1)$ gauge theory of $CP^1$ field coupled to
a massless scalar field.

The Lagrangian gives the following equations of motion
\begin{gather}
\pd^2 \om -\sin\om (\pd_\mu \hn)^2
+\dfrac{\alpha}{8 \kappa^2} \sin\om (\pd_\mu \om
\pd_\nu \hn -\pd_\nu \om \pd_\mu \hn)^2  \nn\\
+\dfrac{\alpha}{\kappa^2} \sin^2 \dfrac{\om}{2}
\pd_\mu \big[ (\pd_\mu \om \pd_\nu \hn
-\pd_\nu \om \pd_\mu \hn) \cdot \pd_\nu \hn \big] \nn\\
- \dfrac{\alpha}{\kappa^2} \sin^2 \dfrac{\om}{2}
\sin\om (\pd_\mu \hn \times \pd_\nu \hn)^2 =0, \nn \\
\pd_\mu \Big\{\sin^2 \dfrac{\om}{2}  \hn \times
\pd_\mu \hn + \dfrac{\alpha}{4\kappa^2} \sin^2 \dfrac{\om}{2}
\big[ (\pd_\nu \om)^2 \hn \times \pd_\mu \hn  \nn\\
-(\pd_\mu \om \pd_\nu \om) \hn \times \pd_\nu \hn \big] \nn\\
+\dfrac{\alpha}{\kappa^2} \sin^4 \dfrac{\om}{2} (\hn \cdot
\pd_\mu \hn \times \pd_\nu \hn) \pd_\nu \hn \Big\}=0.
\label{skeq1}
\end{gather}
Notice that the second equation can be interpreted as
the conservation of $SU(2)$ current originating from
the global $SU(2)$ symmetry of the theory.

It has two interesting limits. First, when
\bea
\om= (2p+1) \pi,
\label{fadd}
\eea
(\ref{skeq1}) is reduced to
\begin{gather}
\hn \times \pd^2 \hn-\dfrac{\alpha}{\kappa^2} 
(\pd_\mu H_\mn) \pd_\nu \hn = 0, \nn\\
H_\mn=\hn \cdot (\pd_\mu \hn \times \pd_\nu \hn)
= \pd_\mu C_\nu - \pd_\nu C_\mu,
\label{sfeq}
\end{gather}
where $C_\mu$ is the magnetic potential of $H_\mn$ defined
by (\ref{mpot}). This is the central equation of Skyrme theory
which allows the monopole, the baby skyrmion, and
the Faddeev-Niemi knot \cite{prl01,plb04,ijmpa08}.

Second, in the spherically symmetric limit
\bea
\om = \om (r),~~~~~\hn = \pm \hat r,
\label{skans}
\eea
(\ref{skeq1}) is reduced to
\bea
&\dfrac{d^2 \om}{dr^2} +\dfrac{2}{r} \dfrac{d\om}{dr}
-\dfrac{2\sin\om}{r^2} +\dfrac{2\alpha}{\kappa^2}
\Big[\dfrac{\sin^2 (\om/2)}{r^2} \dfrac{d^2 \om}{dr^2}  \nn\\
&+\dfrac{\sin\om}{4 r^2} (\dfrac{d\om}{dr})^2
-\dfrac{\sin\om \sin^2 (\om /2)}{r^4} \Big] =0.
\label{skeq2}
\eea
This is the equation used by Skyrme to find the original
skyrmion. Imposing the boundary
condition
\bea
\om(0)= 2\pi,~~~~~\om(\infty)= 0,
\label{skbc}
\eea
we have the skyrmion solution which carries the unit
baryon number \cite{sky}
\bea
&B=-\dfrac{1}{24\pi^2} \Int
\epsilon_{ijk} ~{\rm tr} ~(L_i L_j L_k) d^3r \nn\\
&= -\dfrac{1}{8\pi^2} \int \epsilon_{ijk} \pd_i \om
\big[\hat r \cdot (\pd_j \hat r \times \pd_k \hat r) \big]
\sin^2 \dfrac{\om}{2} d^3r \nn\\
&=1,
\label{bn}
\eea
which represents the non-trivial homotopy $\pi_3(S^3)$
defined by $U$.

The two limits lead us to very interesting physics.
To understand the physical meaning of the first limit
notice that, with (\ref{fadd}) the Skyrme Lagrangian
reduces to the Skyrme-Faddeev Lagrangian
\bea
{\cal L} \rightarrow -\dfrac{\kappa^2}{2} (\pd_\mu \hn)^2
-\dfrac{\alpha}{4}(\pd_\mu \hn \times \pd_\nu \hn)^2,
\label{sflag}
\eea
whose equation of motion is given by (\ref{sfeq}).
This tells that the Skyrme-Faddeev theory becomes
a self-consistent truncation of the Skyrme theory.
This assures that the Skyrme-Faddeev theory is
an essential ingredient (the backbone) of the Skyrme
theory which describes the core dynamics of Skyrme
theory \cite{prl01,plb04,ijmpa08}.

A remarkable feature of the Skyrme-Faddeev
theory is that it can be viewed as a theory of
monopole. In fact, (\ref{sfeq}) has the singular 
monopole solution \cite{prl01,plb04,ijmpa08}
\bea
\hn = \pm \hat r.
\label{mono}
\eea
which carries the magnetic charge
\bea
Q_m = \dfrac{\pm 1}{8\pi} \int \epsilon_{ijk} 
\big[\hat r\cdot (\pd_i \hat r 
\times \pd_j \hat r)\big] d\sigma_k= \pm 1,
\eea
which represents the homotopy $\pi_2(S^2)$ defined
by $\hn$. Of course, (\ref{mono}) has a point singularity 
at the origin which makes the energy divergent. But 
we can easily regularize the singularity with a non-trivial 
$\om$, with the boundary condition (\ref{skbc}). And 
the regularized monopole becomes nothing but 
the well known skyrmion \cite{prl01,plb04,ijmpa08}. 

Moreover, it has the (helical) magnetic vortex solution
made of the monopole-antimonopole pair infinitely
separated apart, and the knot solution which can be
viewed as the twisted magnetic vortex ring made of
the helical vortex whose periodic ends are connected
together. So the monopole plays an essential role in
all these solutions. This shows that the Skyrme-Faddeev
theory, and by implication the Skyrme theory itself,
can be viewed as a theory of monopole.

But perhaps a most important point of the Skyrme-Faddeev
Lagrangian is that it provides a ``missing" link between
Skyrme theory and QCD, because the Skyrme-Faddeev
Lagrangian can actually be derived from
QCD \cite{prl01,plb04,ijmpa08}.

\section{Skyrme Theory and QCD: A Theory of Monopole}

To reveal the deep connection between  Skyrme theory
and QCD, consider the SU(2) QCD
\bea
{\cal L}_{QCD}=-\dfrac14 \vec F_\mn^2.
\label{mqcd}
\eea
Now, we can make the Abelian projection choosing the Abelian
direction to be $\hn$ and imposing the Abelian isometry
\bea
D_\mu \hn=0,
\eea
to the gauge potential. With this we obtain the restricted
potential $\hat A_\mu$ which describes the color neutral
binding gluon (the neuron) \cite{prd80,prl81}
\begin{gather}
\A_\mu\rightarrow \hat A_\mu={\cal A}_\mu
+{\cal C}_\mu,   \nn\\
{\cal A}_\mu=A_\mu \hn,
~~~{\cal C}_\mu=-\dfrac1g \hn \times \pd_\mu \hn.
\label{ap}
\end{gather}
The restricted potential is made of two parts, the naive
Abelian (Maxwellian) part ${\cal A}_\mu$ and the topological
monopole (Diracian) part ${\cal C}_\mu$. Moreover, it has
the full non-Abelian gauge freedom although the holonomy
group of the restricted potential is Abelian.

So we can construct the restricted QCD (RCD) made of
the restricted potential which has the full SU(2) gauge
freedom \cite{prd80,prl81}
\begin{gather}
{\cal L}_{RCD} = -\dfrac{1}{4} \hat F^2_\mn
=-\dfrac{1}{4} (F_\mn-\dfrac1g H_\mn)^2  \nn\\
=-\dfrac{1}{4} F_\mn^2+\dfrac1{2g} F_\mn \hn 
\cdot (\pro_\mu \hn \times \pro_\nu \hn)  
-\dfrac1{4g^2} (\pro_\mu \hn \times \pro_\nu \hn)^2,  \nn\\
F_\mn=\pd_\mu A_\nu-\pd_\nu A_\mu,
~~~H_\mn=\pd_\mu C_\nu-\pd_\nu C_\mu.
\label{rcd}
\end{gather}
This shows that RCD is a dual gauge theory made of 
two Abelian gauge potentials, the electric $A_\mu$
and magnetic $C_\mu$. Nevertheless it has the full 
non-Abelian gauge symmetry.  

Moreover, we can recover the full SU(2) QCD with 
the Abelian decomposition which decomposes 
the gluons to the color neutral neuron and colored 
chromon gauge independently \cite{prd80,prl81}
\begin{gather}
\A_\mu =\hat A_\mu +\vec X_\mu,
~~~\hn \cdot \vec X_\mu=0,   \nn\\
\vec F_\mn=\hat F_\mn+ \hD _\mu \X_\nu 
- \hD_\nu \X_\mu + g\X_\mu \times \X_\nu,
\label{adec}
\end{gather}
where $\vec X_\mu$ describes the gauge covariant 
colored chromon. With this we have the Abelian 
decomposition of QCD 
\begin{gather}
{\cal L}_{QCD} = -\dfrac{1}{4} \vec F^2_\mn
=-\dfrac{1}{4}\hat F_\mn^2-\dfrac{1}{4}(\hD_\mu\X_\nu
-\hD_\nu\X_\mu)^2 \nn\\
-\dfrac{g}{2} {\hat F}_{\mu\nu} \cdot (\X_\mu \times \X_\nu)
-\dfrac{g^2}{4} (\X_\mu \times \X_\nu)^2. 
\label{ecd} 
\end{gather}
This confirms that QCD can be viewed as RCD made of 
the binding gluon, which has the colored valence gluon 
as its source \cite{prd80,prl81}. 

Now we can reveal the connection between the Skyrme 
theory and QCD. Let us start from RCD and assume that
the binding gluon (restricted potential) acquires a mass
term after the confinement. In this case (\ref{rcd}) becomes
\bea
&{\cal L}_{RCD}\rightarrow  -\dfrac{1}{4} \hat F^2_\mn
-\dfrac{m^2}{2} \hat A_\mu^2  \nn\\
&=-\dfrac{1}{4} F_\mn^2
+\dfrac1{2g} F_\mn \hn \cdot (\pd_\mu \hn \times \pd_\nu \hn)  \nn\\
&-\dfrac1{4g^2} (\pd_\mu \hn \times \pd_\nu \hn)^2
-\dfrac{m^2}{2} \big[A_\mu^2+ (\pd_\mu \hn)^2\big].
\label{mrcd}
\eea
Of course, the mass term breaks the gauge symmetry, 
but this can be justified because the binding gluons 
could acquire mass after the confinement sets in.

Integrating out the $A_\mu$ potential (or simply putting
$A_\mu=0$), we can reduce (\ref{mrcd}) to
\bea
&{\cal L}_{RCD}\rightarrow-\dfrac1{4g^2}
(\pd_\mu \hn \times \pd_\nu \hn)^2
-\dfrac{m^2}{2} (\pd_\mu \hn)^2,
\label{mrcd1}
\eea
which becomes nothing but the Skyrme-Faddeev 
Lagrangian (\ref{sflag}) when $\kappa^2=m^2$ 
and $\alpha=1/g^2$. This tells that the Skyrme-Faddeev 
Lagrangian can actually be derived from RCD. And this 
is a mathematical derivation. This confirms that the Skyrme
theory and QCD is closely related, more closely than it
appears.

This has deep consequences. Notice that $\hn$ which
provides the Abelian projection in QCD naturally
represents the monopole topology $\pi_2(S^2)$, and
can describe the monopole. In fact, it is well known
that $\hn=\hat r$ becomes exactly the Wu-Yang
monopole solution \cite{prd80,prl81,prl80}. On
the other hand the above exercise tells that this $\hn$
is nothing but the normalized pion field in the Skyrme
theory. This strongly implies that $\hn=\hat r$ can
also describe the monopole solution in the Skyrme
theory.  Indeed, (\ref{mono}) is precisely the Wu-Yang
monopole transplanted in the Skyrme theory.

What is more, the Skyrme theory has more complicated
monopole solutions. With $\hn=\hat r$ and the boundary
condition
\bea
\omega(0)=\pi,~~~~\omega(\infty)=0,~~~~({\rm mod}~2\pi)
\label{+mbc}
\eea
we can find a monopole solution which reduces to the singular
solution (\ref{mono}) near the origin. The only difference
between this and (\ref{mono}) is the non-trivial dressing of
the scalar field $\omega$, so that it could be interpreted as
a dressed monopole. This dressing, however, is only partial
because this makes the energy finite at the infinity, but not
at the origin.

Similarly, with the following boundary condition
\bea
\omega(0)=2\pi,~~~~\omega(\infty)=\pi,~~~~({\rm mod}~2\pi)
\label{-mbc}
\eea
we can obtain another monopole and half-skyrmion solution
which approaches the singular solution near the infinity.
Here again the partial dressing makes the energy finite at
the origin, but not at the infinity. So the partially dressed
monopoles still carry an infinite energy.

\begin{figure}
\begin{center}
\includegraphics[height=4.5cm, width=7cm]{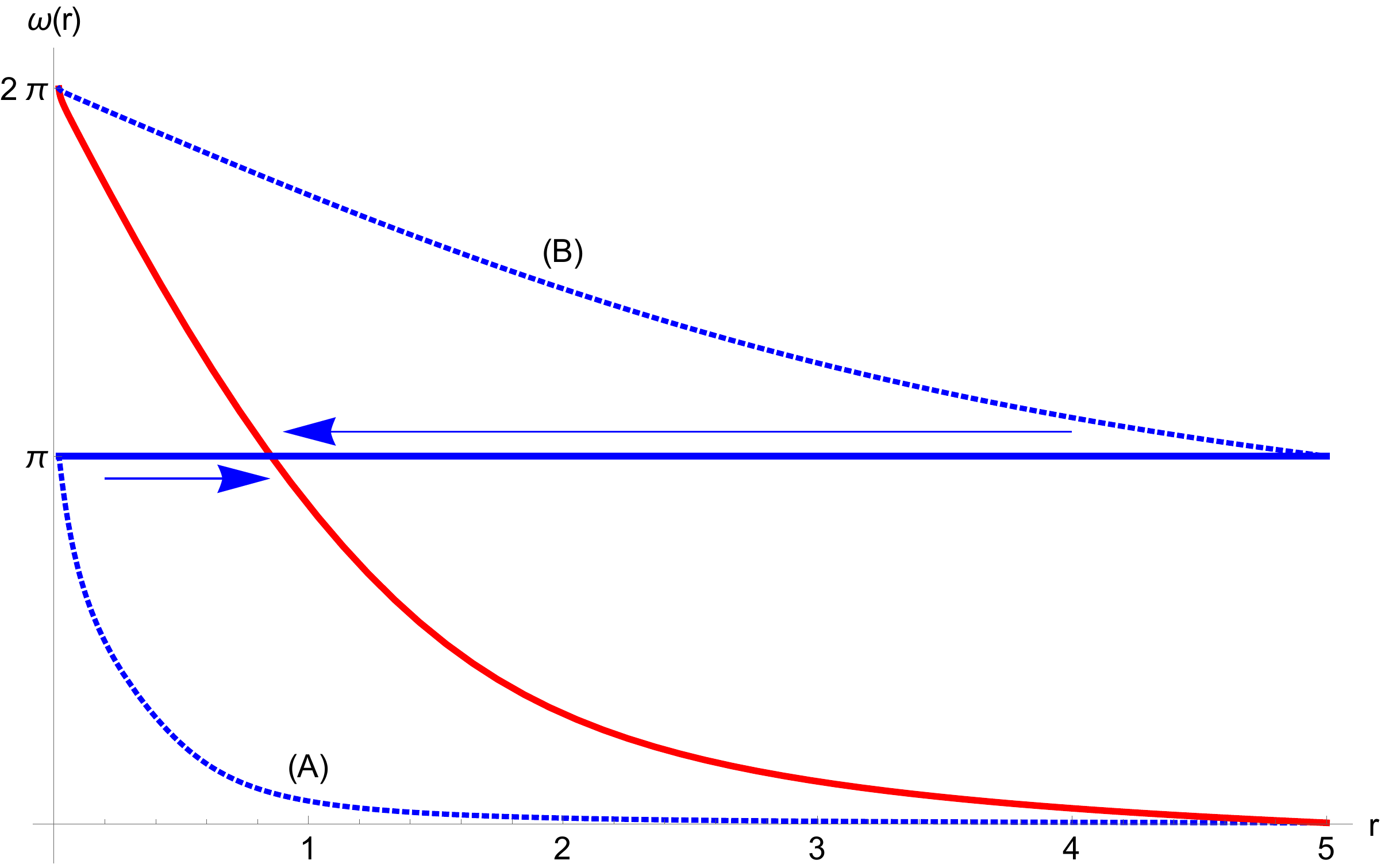}
\caption{\label{mosky} The singular Wu-Yang monopole 
(the blue line), the two half skyrmions (A) and (B) as 
partially dressed singular monopoles (blue curves), 
and the regular skyrmion (red curve) obtained combining 
the two half skyrmions.}
\end{center}
\end{figure}

Clearly these solutions carry the unit magnetic charge of
the homotopy $\pi_2(S^2)$ defined by $\hn$ \cite{ijmpa08}
\bea
&M = \dfrac{\pm 1}{8\pi} \int \epsilon_{ijk} \big[\hat r
\cdot (\pd_i \hat r \times \pd_j \hat r)\big] d\sigma_k
= \pm 1,
\label{mone}
\eea
but carry a half baryon number
\bea
&B= -\dfrac{1}{8\pi^2} \int \epsilon_{ijk} \pd_i \om
\big[\hat r \cdot (\pd_j \hat r \times \pd_k \hat r) \big]
\sin^2 \dfrac{\om}{2} d^3r \nn\\
&=-\dfrac{1}{\pi} \int \sin^2 \dfrac{\om}{2} d\om=\dfrac12.
\label{bnhalf}
\eea
This is due to the boundary conditions (\ref{+mbc}) and
(\ref{-mbc}). So it describes a half-skyrmion.

The half-skyrmions have two remarkable features.
First, combining the two half-skyrmions we can
form a finite energy soliton, the fully dressed
skyrmion \cite{ijmpa08}. This must be clear because,
putting the boundary conditions (\ref{+mbc})
and (\ref{-mbc}) together we recover the boundary
condition (\ref{skbc}) of the skyrmion. So putting
the two partially dressed monopoles together we
obtain the well known finite energy skyrmion solution.

This is summarized in Fig. \ref{mosky}, where the blue line
represents the singular monopole, the dotted curves (A) and
(B) represent the singular half skyrmions, and the red curve
represents the regular skyrmion. Notice that in all these
solutions we have $\hn=\hr$, so that $M=1$. This confirms
that the skyrmion is nothing but the Wu-Yang monopole
of QCD, transplanted and regularized to have a finite
energy by the massless scalar field $\om$ in the Skyrme
theory \cite{prl01,plb04,ijmpa08}.

The other remarkable feature of the half-skyrmions
is that the monopole number (\ref{mone}) and
the baryon number (\ref{bnhalf}) are different.
This is very interesting because, in skyrmions
the baryon number is commonly identified by
the rational map which defines the monopole
number. This suggests that the baryon number 
and the monopole number are one and the same
thing \cite{houg}. But obviously this is not true
for the half-skyrmions. In the following we will 
argue that this need not be ture, and show that 
the skyrmions in general have two different topological 
numbers.

\begin{figure}
\begin{center}
\includegraphics[width=5cm, height=5cm]{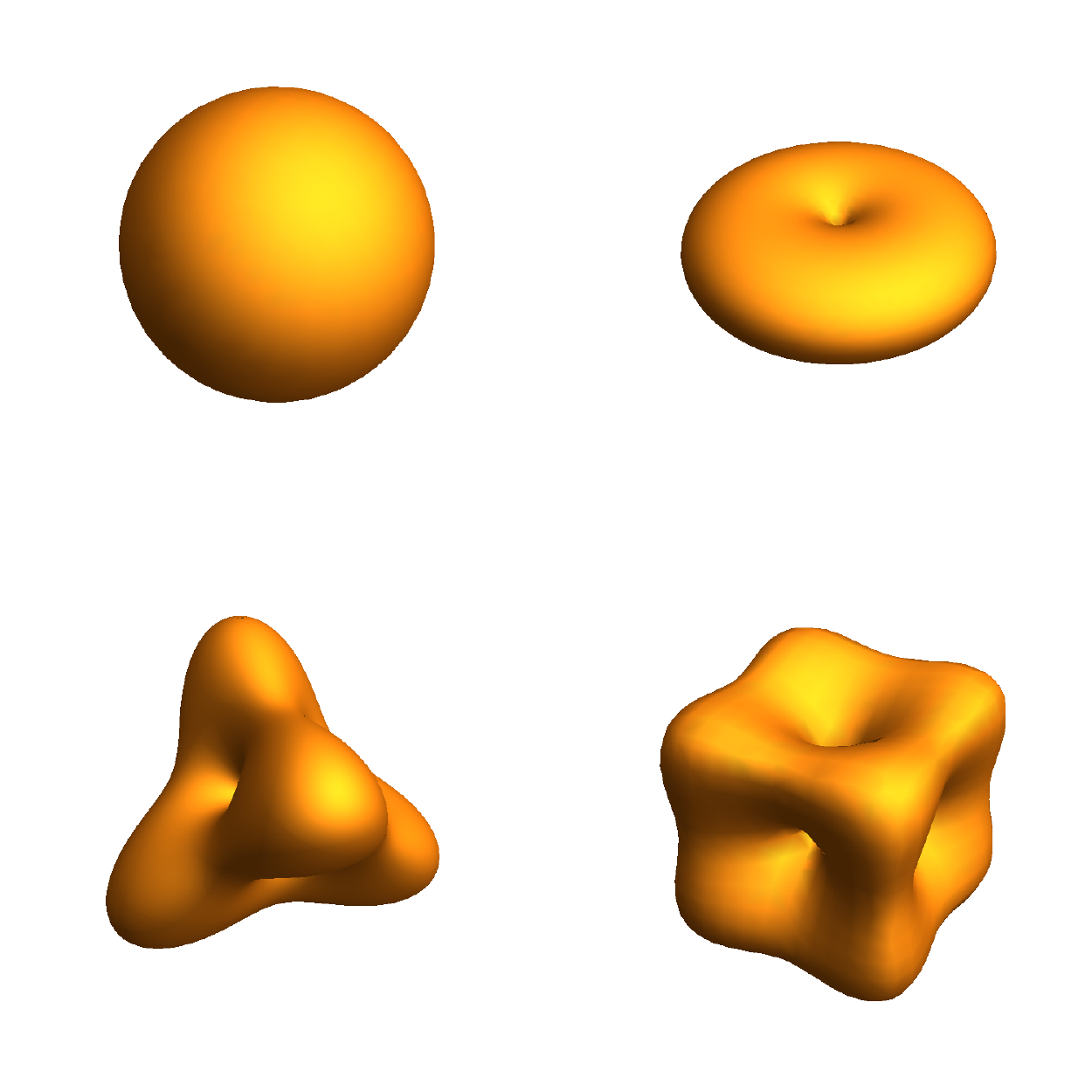}
\caption{\label{oldsoln} The well known (non spherically
symmetric) numerical multi-skyrmion solutions with
baryon number 1,2,3,4.}
\end{center}
\end{figure}

\section{A New Classification of Skyrmion}

The above argument clearly shows that the skyrmion has
the monopole topology. And this monopole topology
reappear in all popular non spherically symmetric
multi-skyrmion solutions \cite{man,bat,houg,sut,piet1,piet2}.
Indeed, it is well known that the monopole topology of
the rational map $\pi_2(S^2)$ given by $\hn$,
\bea
&M= \dfrac{1}{8\pi} \int \epsilon_{ijk} \big[\hn
\cdot (\pd_i \hn \times \pd_j \hn)\big] d\sigma_k
= m,
\label{mnm}
\eea
together with the boundary condition (\ref{skbc}),
has played the fundamental role for many people
to construct these multi-skyrmion solutions. Some
of these multi-skyrmion solutions are shown in 
Fig. \ref{oldsoln}.

On the other hand, these solutions have always been
interpreted to have the baryon topology $\pi_3(S^3)$
which have the baryon number given by
\bea
&B= -\dfrac{1}{8\pi^2} \int \epsilon_{ijk} \pd_i \om
\big[\hn \cdot (\pd_j \hn \times \pd_k \hn) \big]
\sin^2 \dfrac{\om}{2} d^3r \nn\\
&= \dfrac{1}{8\pi} \int \epsilon_{ijk} \big[\hn
\cdot (\pd_i \hn \times \pd_j \hn)\big] d\sigma_k
=m.
\label{bnm}
\eea
But obviously this baryon number is precisely the monopole
number shown in (\ref{mnm}), so that these solutions
have $B=M$. Nevertheless, the monopole topology
$\pi_2(S^2)$ and the baryon topology $\pi_3(S^3)$ is
clearly different. If so, we may ask what is the relation
between the monopole topology and the baryon topology.

\begin{figure}
\begin{center}
\includegraphics[height=5cm, width=7cm]{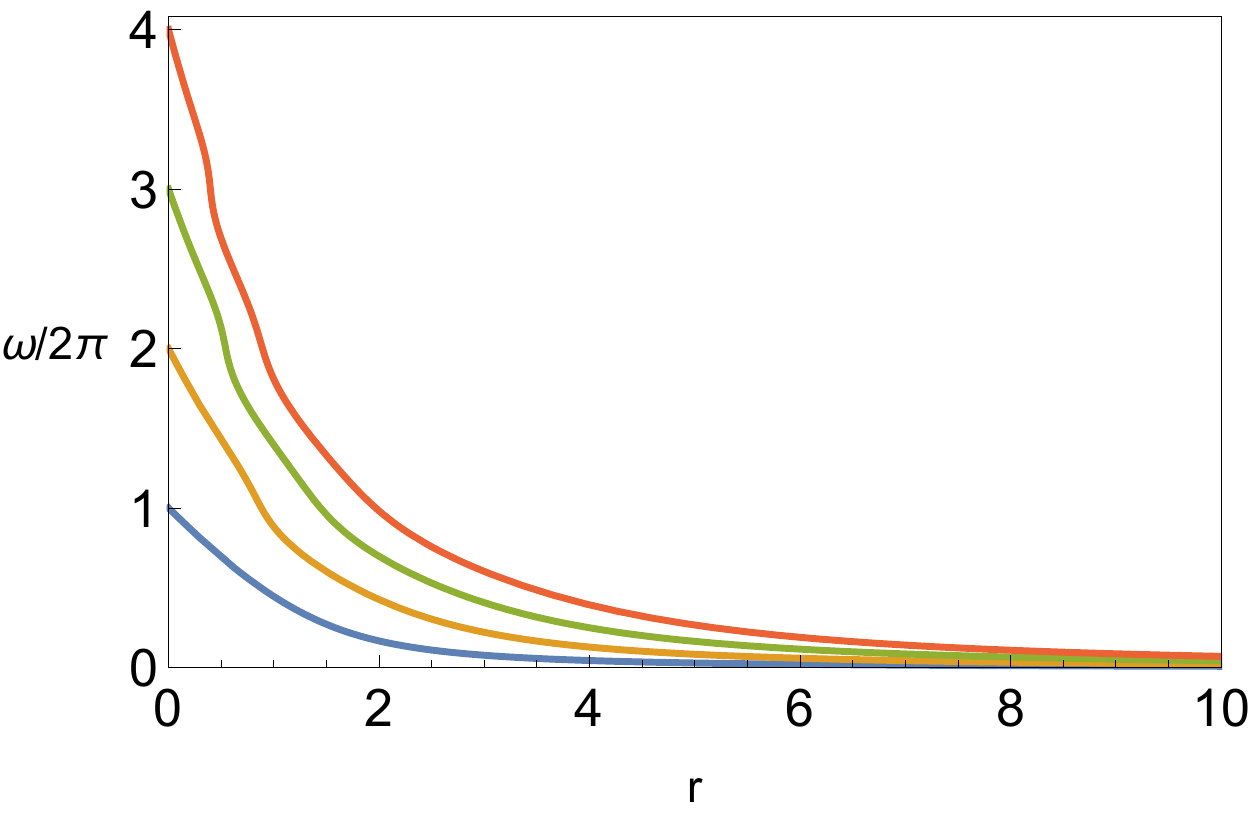}
\caption{\label{skysoln} The spherically symmetric skyrmions
with baryon number 1,2,3,4, which should be contrasted
with the solutions shown in Fig. \ref{oldsoln}.}
\end{center}
\end{figure}

The Skyrme equation (\ref{skeq2}) in the spherically
symmetric limit plays an important role to clarify this
point. To see this notice that, although the SU(2)
matrix $U$ is periodic in $\om$ variable by $4\pi$,
$\om$ itself can take any value from $-\infty$ to
$+\infty$. So we can obtain the spherically symmetric
multi-skyrmion solutions generalizing the boundary
condition (\ref{skbc}) to \cite{sky,witt,rho,bogo,prep}
\bea
\omega(0)= 2\pi n,~~~~~\omega(\infty)= 0,
\label{skbcn}
\eea
with an arbitrary integer $n$. Some of the spherically
symmetric multi-skyrmion solutions are shown in
Fig. \ref{skysoln}. These are, of course, the solutions
that Skyrme originally proposed to identify as the nuclei
with baryon number larger than one \cite{sky,rho}. But
soon after they are dismissed as uninteresting because
they have too much energy.

An interesting features of the spherically symmetric
solutions is that whenever the curve passes through
the values $\om=2\pi n$, it become a bit steeper.
This is because (as we will see soon) these points are
the vacua of the theory, and the steep slopes shows
that the energy likes to be concentrated around
these vacua.

\begin{figure}
\begin{center}
\includegraphics[height=5cm, width=7cm]{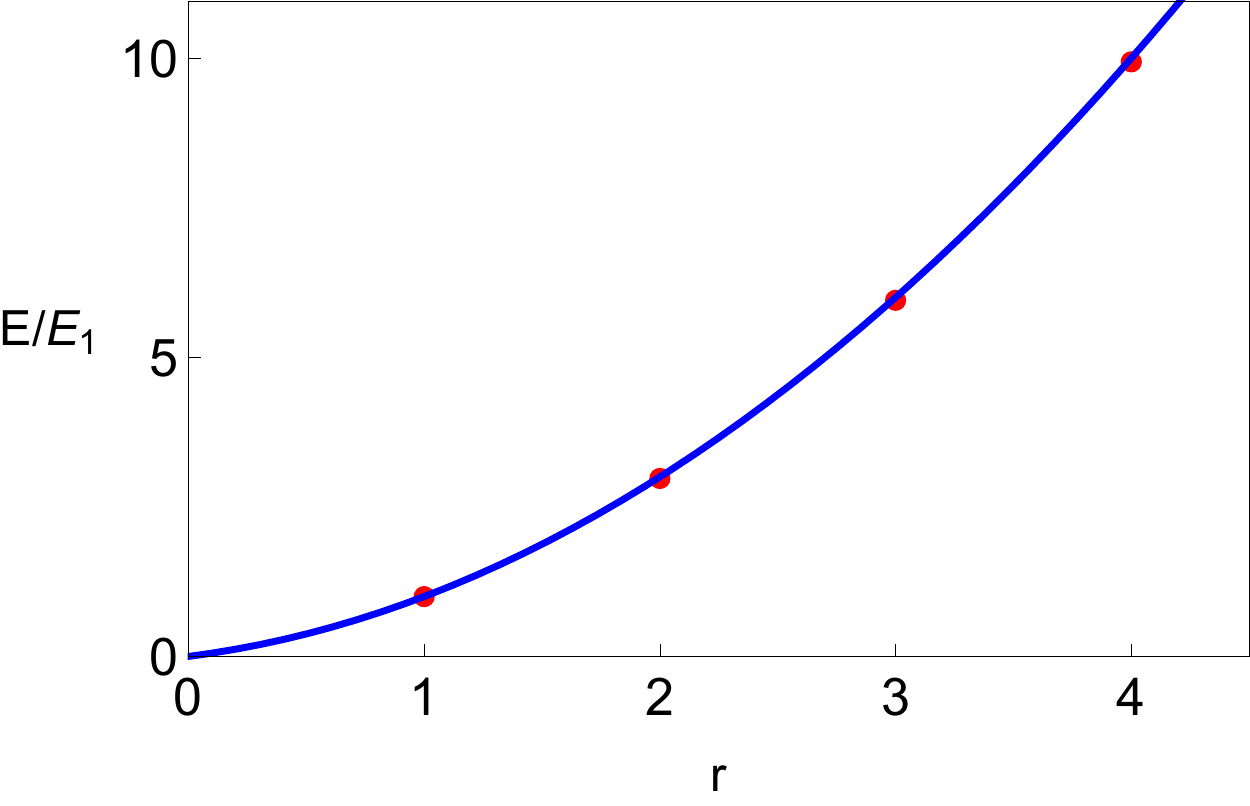}
\caption{\label{Enn} The energy of the spherically symmetric
solutions with baryon number 1,2,3,4. The numerical
fit (the blue curve) and the $n(n+1) E_1/2$ curve (the green
curve) are almost indistinguishable.}
\end{center}
\end{figure}

Another interesting feature of the spherically symmetric
skyrmions is the energy, which is given by \cite{ijmpa08}
\bea
&E =\dfrac{\pi \kappa^2}{2}  \Int_0^\infty \Big\{\Big(r^2
+\dfrac{2\alpha}{\kappa^2} \sin^2\dfrac{\om}{2}\Big)
\Big(\dfrac{d\om}{dr}\Big)^2  \nn\\
&+8 \Big(1+\dfrac{\alpha}{2\kappa^2 r^2} \sin^2
\dfrac{\om}{2} \Big) \sin^2 \dfrac{\om}{2} \Big\} dr  \nn\\
&= \pi {\sqrt \alpha} \kappa \Int^{\infty}_{0}
\Big[x^2 \left(\dfrac{d\om}{dx}\right)^2
+ 8 \sin^2{\dfrac{\om}{2}} \Big] dx,
\label{sken}
\eea
where $x=\kappa r/{\sqrt \alpha}$. From this we can easily
calculate their energy. This is shown in Fig. \ref{Enn}.

Numerically the baryon number dependence of the energy
is given by \cite{rho,bogo,prep}
\bea
E_n\simeq \dfrac{n(n+1)}{2} E_1.
\label{en}
\eea
This has two remarkable points. First, the baryon number
dependence of the energy is quadratic. This, of course,
means that the energy of the skyrmion with baryon number
$n$ is bigger than the sum of $n$ lowest energy skyrmion.
So the radially excited skyrmions are unstable. This is
not the case in the popular multi-skyrmion solutions shown
in Fig. \ref{oldsoln}. They have positive binding energy,
so that the energy of the skyrmion with baryon number $B$
becomes smaller than the $B$ sum of the lowest energy
skyrmion. This was the main reason why the spherically
symmetric solutions have not been considered seriously
as the model of heavy nuclei.

But actually (\ref{en}) makes the spherically symmetric
solutions mathematically more interesting, because
(\ref{en}) is almost exact. In fact Fig. \ref{Enn} shows
that the mathematical curve $E_n= n(n+1)E_1/2$ and
the numerical fit is almost indistinguishable. Intuitively
this could be understood as follows. Roughly speaking,
the kinetic energy (first part) and the potential energy
(second part) of (\ref{sken}) become proportional to
$n^2$ and $n$, and the two terms have an equal
contribution due to the equipartition of energy. But
the truth is more complicated than this, and we certainly
need a mathematical explanation of (\ref{en}).

\begin{figure}
\begin{center}
\includegraphics[height=5cm, width=7cm]{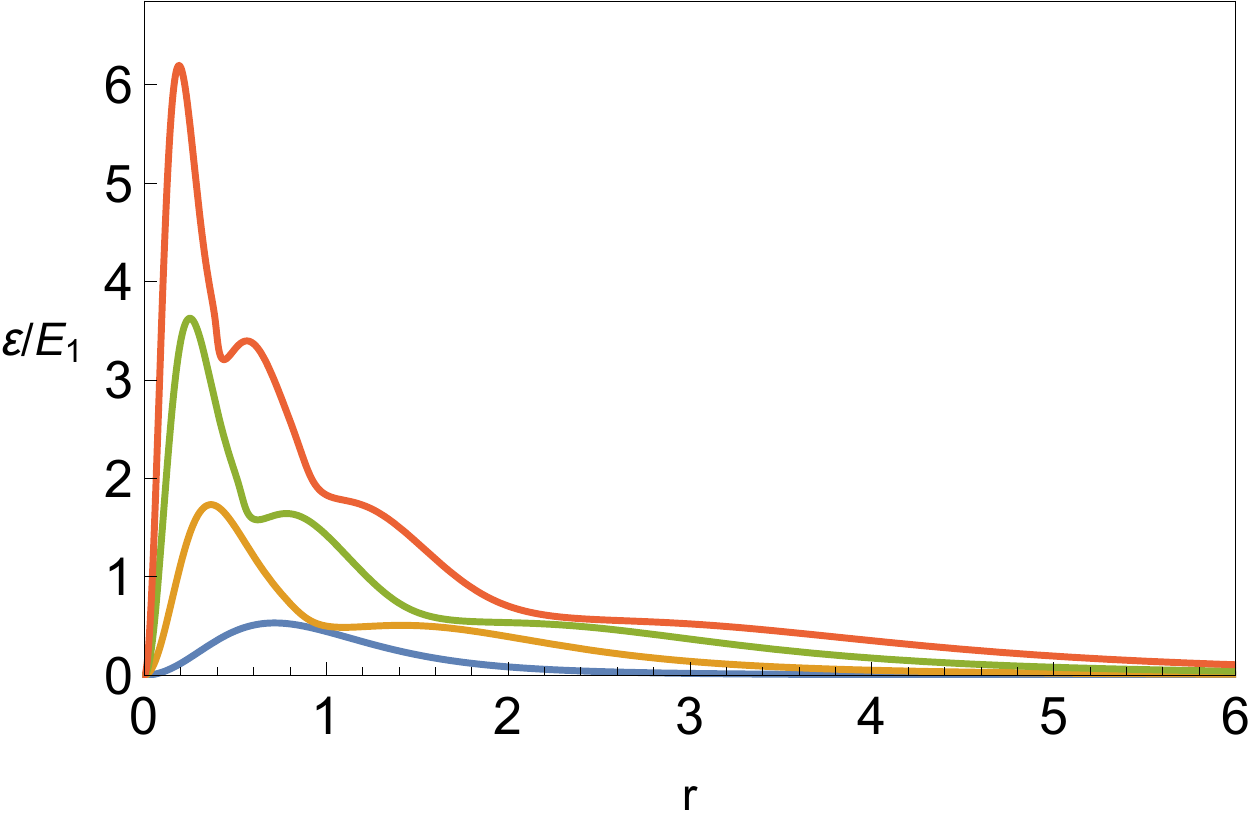}
\caption{\label{edenn} The radial energy density of
the spherically symmetric skyrmions with baryon number
1,2,3,4. The density functions are normalized
to make the integral of the $B=1$ solution to be
the unit.}
\end{center}
\end{figure}

The energy density of the solutions is shown in Fig.
\ref{edenn}. Clearly the $B=n$ solution has $n$ local
maxima, which indicates that it is made of n shells
of unit skyrmions. Moreover, as we have remarked
the energy density has the local maxima at $\om=2\pi n$.
So they describe the shell model of nuclei, made of
$n$ shells located at $\om=2\pi n$. This tells that
the spherically symmetric solutions could be viewed
as radially extended skyrmions of the original skyrmion.
In this interpretation the baryon number $n$ of these
skyrmions can be identified as the radial, or more
properly the shell, number \cite{piet1,piet2}. And this
baryon number is fixed by the winding number
$\pi_1(S^1)$ of the angular variable $\om$.

The contrast between the popular non spherically
symmetric solutions shown in Fi‍g. \ref{oldsoln}
and the sphrically symmetric solutions shown in
Fig. \ref{skysoln} is unmistakable. But the contrast
is not just in the appearance. They are fundamentally
different. In particular, the spherically symmetric
solutions have a very important implication.

To understand this, notice that these solutions have
the monopole number given by
\bea
&M= \dfrac{1}{8\pi} \int \epsilon_{ijk} \big[\hr
\cdot (\pd_i \hr \times \pd_j \hr)\big] d\sigma_k =1.
\label{mn1}
\eea
On the other hand, their baryon number is given by 
the winding number $\pi_1(S^1)$ of $\om$ determined 
by the boundary condition (\ref{skbcn}),
\bea
&B= -\dfrac{1}{8\pi^2} \int \epsilon_{ijk} \pd_i \om
\big[\hat r \cdot (\pd_j \hat r \times \pd_k \hat r) \big]
\sin^2 \dfrac{\om}{2} d^3r \nn\\
&=-\dfrac{1}{\pi} \int \sin^2 \dfrac{\om}{2} d\om=n.
\label{bnn}
\eea
So, unlike the popular non spherically symmetric
multi-skyrmions shown in Fig. \ref{oldsoln}, the baryon
number and the monopole number of these solutions
are different. Moreover, the baryon number is not given
by the rational map, but by the winding number.

This confirms that the skyrmions actually do carry two
topological numbers, the baryon number of $\pi_3(S^3)$
and the monopole number of $\pi_2(S^2)$ \cite{epjc17}.
And the spherically symmetric solutions play the crucial
role to demonstrate this.

In this scheme the skyrmions are classified by
$(m,b)$, by the monopole number $m$ and
the baryon number $b$. And the popular (non
spherically symmetric) solutions become the $(m,m)$
skyrmions and the spherically symmetric solutions
become the $(1,n)$ skyrmions.

This has a deep consequence. Based on this we
can actually show that the baryon number can be
decomposed to the monopole number and the radial
(shell) number and expressed by the product of
the two numbers \cite{epjc17}. To show this we
discuss the characteristic features of the spherically
symmetric solutions first.

This implies that we could replace the baryon number
with the shell number, and classify the skyrmions by
the monopole number $m$ and shell number $n$ by
$(m,n)$, instead of $(m,b)$. In other words we could
also classify the skyrmions by the monopole topology
$\pi_2(S^2)$ of $\hn$ and the U(1) (i.e., shell) topology
$\pi_1(S^1)$ of $\om$. In this scheme the baryon number
of the $(m,n)$ skyrmion is given by $B=mn$. This must be
clear from (\ref{bn}), which tells that the baryon number
is made of two parts, the $\pi_2(S^2)$ of $\hn$ and
$\pi_1(S^1)$ of $\om$.

According to this classification the popular (non spherically
symmetric) skyrmions become the $(m,1)$ skyrmions,
the radially extended spherically symmetric skyrmions
become the $(1,n)$ skyrmions. The justification for this
classification comes from the following observation. First,
the $S^3$ space (both the real space and the target space)
in the baryon topology $\pi_3(S^3)$ admits the Hopf
fibering $S^3\simeq S^2\times S^1$. Second, the two
variables $\hn$ and $\om$ of the Skyrme theory naturally
accommodate the monopole topology $\pi_2(S^2)$ and
the shell topology $\pi_1(S^1)$.

Furthermore, in this classification the baryon number
is given by the product of the monopole number and
the shell number, or $B=mn$. To see this notice that
the Hopf fibering of $S^3$ has an important property
that locally it can be viewed as a Cartesian product of 
$S^2$ and $S^1$. So we have
\bea
&B =-\dfrac{1}{8\pi^2} \Int \epsilon_{ijk} \pd_i \om
\big[\hn \cdot (\pd_j \hn \times \pd_k \hn) \big]
\sin^2 \dfrac{\om}{2} d^3r \nn\\
&=-\dfrac{1}{8\pi^2} \Int \pd_i \om
\big[\hn \cdot (\pd_j \hn \times \pd_k \hn) \big]
\sin^2 \dfrac{\om}{2} dx^i \wedge dx^j \wedge dx^k \nn\\
&=-\dfrac{1}{8\pi^2} \Int \sin^2 \dfrac{\om}{2} d\om
\times \epsilon_{ijk} \big[\hn \cdot (\pd_i \hn
\times \pd_j \hn) \big] d\Sigma_k \nn\\
&=\dfrac{n}{8\pi} \Int \epsilon_{ijk} \big[\hn \cdot
(\pd_i \hn \times \pd_j \hn) \big] d\Sigma_k=mn,
\label{bmn}
\eea
where $d\Sigma_k=\epsilon_{ijk} dx^i \wedge dx^j/2$.
Clearly the last integral is topologically equivalent
to (\ref{mono}), which assures the last equality.
This shows that the baryon number of the skyrmion
is given by the product of the monopole number and
the shell number. Obviously both $(m,1)$ and $(1,n)$
skyrmions are the particular examples of this.

Another way to show this is to introduce a generalized
coordinates $(\eta,\alpha,\beta)$ for $R^3$ whose line
element is given by \cite{fab}
\begin{gather}
ds^2= f(\eta) d \eta^2 +d \Sigma^2(\alpha,\beta),
\end{gather}
in which $\eta$ represents the radial coordinate $R^1$
and $(\alpha,\beta)$ represents the $S^2$ such that
$\om=\om(\eta)$ and $\hn=\hn(\alpha,\beta)$. Again this
is possible because the hopf fibering of $S^3$ can be
expressed by a locally Cartesian product of $S^2$ and
$S^1$. In this coordinates we clearly have
\begin{gather}
B =-\dfrac{1}{8\pi^2} \Int \epsilon_{ijk} \pd_i \om
\big[\hn \cdot (\pd_j \hn \times \pd_k \hn) \big]
\sin^2 \dfrac{\om}{2} d^3x \nn\\
=-\dfrac{1}{8\pi^2} \Int \sin^2 \dfrac{\om}{2} \frac{d\om}{d\eta}  \nn\\
\times 2 \Int \big[\hn \cdot (\frac{\pd \hn}{\pd \alpha}
\times \frac{\pd \hn}{\pd \beta}) \big] d\alpha \wedge d\beta \nn\\
=-\dfrac{1}{4\pi^2} \Int \sin^2 \frac{\om}{2} d \om
\Int \big[\hn \cdot (d \hn \wedge d\hn) \big]  \nn\\
=mn.
\label{bnmn}
\end{gather}
So in this local direct product coordinates, it is
straightforward to prove $B=mn$. Notice that
the third equality telles that the baryon number
can be expressed in a coodinate independent form.

Clearly the spherically symmetric solutions satisfy
this criterion. In this case, $\pi_3(S^3)$ naturally
decomposes to $\pi_2(S^2)$ of $\hn$ and $\pi_1(S^1)$
of $\om$, so that the baryon number $B$ is decomposed
to the $S^2$ wrapping number $m$ and the $S^1$
winding number $n$. But the spherical symmetry
requires $\hn=\hat r$, and restricts $m=1$.

Again this follows from two facts. First, the fact that
the Hopf fibering allows $S^3$ to decompose $S^2 \times S^1$,
and the fact that in Skurme theory $\om$ and $\hn$ naturally
decompose the target space $S^3$ to $S^1$ and $S^2$. With
this we have $\pi_3(S^3) \simeq \pi_2(S^2) \times \pi_1(S^1)$.
This strongly support that the baryon topology $\pi_3(S^3)$
can be decomposed to $\pi_2(S^2)$ of $\hn$ and $\pi_1(S^1)$
of $\om$.

If so, one might wonder whether the popular skyrmions
can also be made to carry two different topological
numbers. This should be possible. To see this remember
that the integer $n$ in the $(1,n)$ skyrmions describes
the radial (shell) number which describes the shell
structure of the spherically symmetric skyrmions. And
this shell structure is provided by the angular variable
$\om$ which allows the radial extension of the original
$(1,1)$ skyrmion. So we can generalize the $(m,1)$
skyrmion to have similar shell structure.

In fact, we may obtain the ``radially extended" solutions
of the non spherically symmetric skyrmions numerically,
generalizing the boundary condition (\ref{skbc}) to
(\ref{skbcn}), requiring \cite{piet1,piet2}
\bea
&\om(r_k)=2\pi k,~~~(k=0,1,2,...n),  \nn\\
&r_0=0 ~\langle~r_1~\langle~...~\langle~r_n=\infty,
\eea
keeping the rational map number $m$ of $\hn$ unchanged.
With this we could find new solutions numerically minimizing
the energy, varying $r_k~(k=1,2,...,n-1)$. This way we can
add the shell structure and the shell number to the $(m,m)$
skyrmion. In this case the baryon number of the radially
extended skyrmions should become $B=mn$.

Obviously the spherically symmetric solutions satisfy
this criterion. In this case, $\pi_3(S^3)$ naturally
decomposes to $\pi_2(S^2)$ of $\hn$ and $\pi_1(S^1)$
of $\om$, so that the baryon number $B$ is decomposed
to the $S^2$ wrapping number $m$ and the $S^1$
winding number $n$. But the spherical symmetry
requires $\hn=\hat r$, and restricts $m=1$.

But the above argument tells that in general (even for non
spherically symmetric skyrmions) we may introduce the
generalized coordinates such that $\hn=\hn(\alpha,\beta)$
and $\om=\om(\gamma)$, and can still make the ``radial"
extension of the skyrmions to obtain the radially extended
$(m,n)$ skyrmions which have $n$ shells numerically,
generalizing the boundary condition (\ref{skbc}) to
(\ref{skbcn}).

Again this becomes possible because the Skyrme theory is
described by two variables $\hn$ and $\om$ which naturally
accommodate the monopole topology $\pi_2(S^2)$ and
the shell topology $\pi_1(S^1)$.  This is very important,
because mathematically there is no way to justify the replacement
of the $\pi_3(S^3)$ topology by two independent $\pi_2(S^2)$
topology and $\pi_1(S^1)$ topology.

Now, one may ask about the stability of the $(m,n)$ skyrmions.
Clearly an $(m,n)$ skyrmion does not have to be stable, and
could decay to lower energy skyrmions as far as the decay
is energetically allowed. For example, (\ref{en}) clearly
tells that the $(1,n)$ skyrmion can decay to lower energy
skyrmions. This is natural. An interesting and important
question here is whether the two topological numbers $m$
and $n$ are conserved independently or not. Certainly
the baryon topology and the monopole topology are mathematically
independent. This means that  the baryon number and
the monopole number must be conserved separately. This
(with $\delta b=n~\delta n+ n~\delta m$) automatically
guarantees that the shell number must also be conserved.

This means that the two numbers $m$ and $n$ must be
conserved separately, so that the $(n,m)$ skyrmion could
decay to $(n_1,m_1)$ and $(n_2,m_2)$ skyrmions when
$n=n_1+n_2$ and $m=m_1+m_2$. But there is no way that
the two topological numbers $m$ and $n$ can be transformed
to each other. This tells that the skyrmions retains
the topological stability of two topology independently,
even when they are classified by two topologial numbers.

The above discussions raise another deep question.
As we have remarked, when $\om=(2n+1)\pi$, 
the Skyrme theory reduces to the Skyrme-Faddeev 
theory. In this limit the Skyrme theory has the knot 
solutions described by $\hn$ whose topology is given 
by $\pi_3(S^2)$ \cite{prl01,plb04,ijmpa08}. And in 
these knots, $\om$ is not activated. If so, one might 
ask if we can dress the knots with $\om$ and add 
the shell structure to the knots to have two topological 
numbers $\pi_1(S^1)$ and $\pi_3(S^2)$. This is a mind 
boggling question which certainly deserves more study.

\section{Baby Skyrmions with Two Topological Numbers}

In this section we present another evidence which 
tells that the skyrmion can be generalized to have 
two topological number, that the baby skyrmion can 
also be gereralized to have two topological numbers.
To do that we first review the old baby skyrmions 
and obtain some new baby skyrmions first. 

Let us choose the cylindrical coordinates $(\varrho,\varphi,z)$ 
and adopt the ansatz \cite{ijmpa08,piet}
\begin{gather}
\hn= \left(\begin{array}{c}  \sin{f(\varrho)}\cos{m \varphi} \\
\sin{f(\varrho)}\sin{m \varphi} \\
\cos{f(\varrho)}
\end{array} \right).
\label{bskyans}
\end{gather}
With this the Skyrme-Faddeev equation (\ref{sfeq}) is 
reduced to
\begin{gather}
\big(1+\dfrac{\alpha}{\kappa^2} \dfrac{m^2}{\varrho^2}
\sin^2{f}\big) \ddot{f}\nn\\
+\dfrac{1}{\varrho} \Big(1
+\dfrac{\alpha}{\kappa^2} \dfrac{m^2}{\varrho}
\dot{f} \sin{f}\cos{f}- \dfrac{\alpha}{\kappa^2}
\dfrac{m^2}{\varrho^2}\sin^2{f} \Big) \dot{f}  \nn\\
- \dfrac{m^2}{\varrho^2} \sin{f}\cos{f}=0.
\label{bskyeq}
\end{gather}
Solving this with the boundary condition
\begin{gather}
f(0)= (2p+1) \pi,~~~f(\infty)=0,
\label{bskybc}
\end{gather}
we obtain the non-Abelian vortex solutions which has
the quantized magnetic flux along the $z$-axis
\begin{gather}
\Phi =\Int H_{\varrho\varphi} d\varrho d\varphi
= 4\pi m,
\label{mqn}
\end{gather}
known as the baby skyrmion \cite{ijmpa08,piet}.

The origin of this quantization of the magnetic flux,
of course, is topological. Clearly $\hn$ defines 
the mapping $\pi_2(S^2)$ from the compactified 
$xy$-plane $S^2$ to the target space $S^2$ of 
SU(2) space defined by $\hn^2=1$ \cite{piet}. And 
this, of course, is precisely the rational mapping 
which defines the monopole topology of the skyrmion. 

\begin{figure}
\begin{center}
\includegraphics[height=5cm, width=9cm]{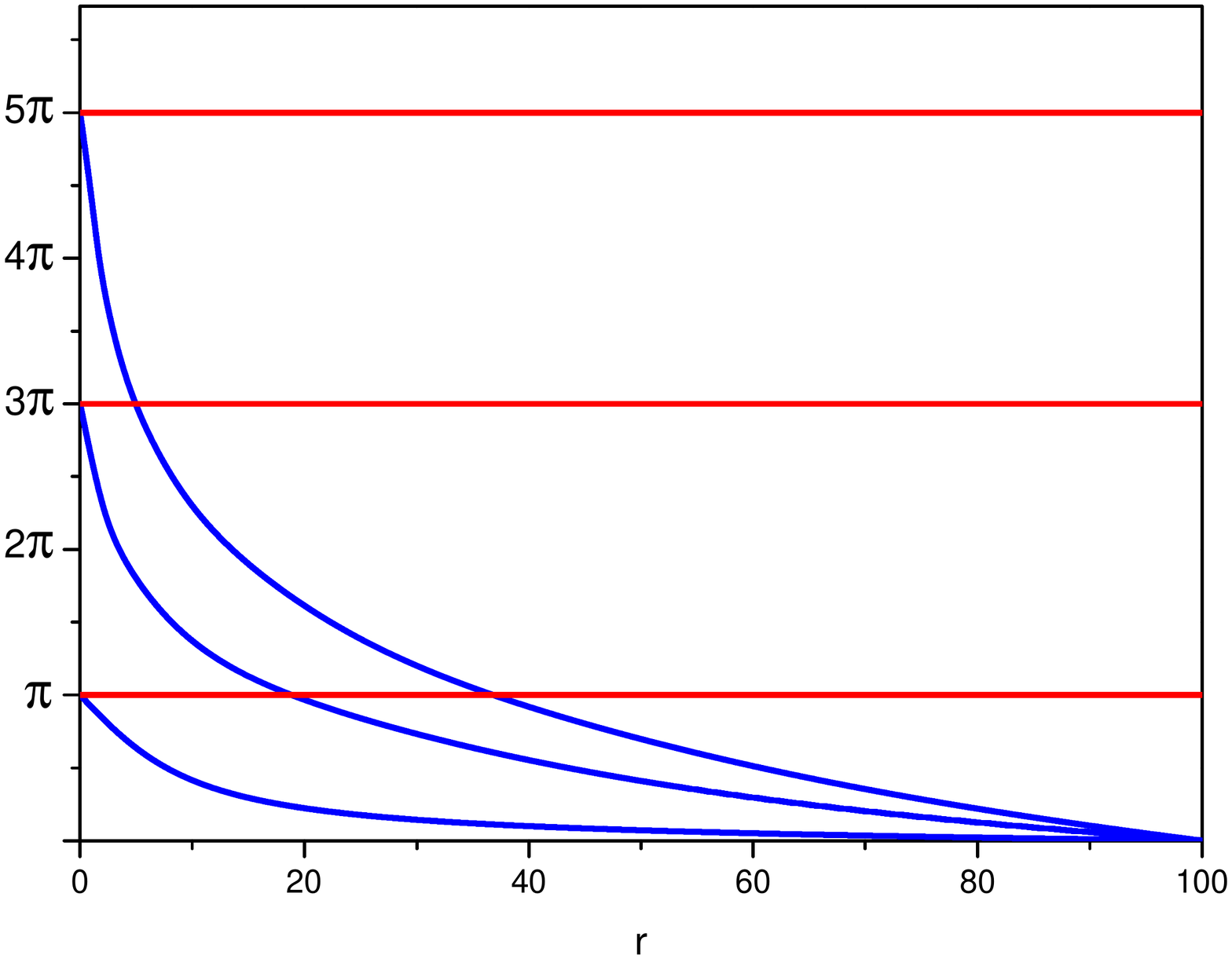}
\caption{\label{cbsky2} The singular and regularized
baby skyrmion solutions which carry the $4\pi m$ 
magnetic flux for $m=1,2,3$. The red straight lines 
represent the singular solutions, and the blue curves 
represent the regularized solutions.}
\end{center}
\end{figure}

But notice that (\ref{bskyeq}) has the singular vortex 
solutions given by
\begin{gather}
f(\varrho)= (2p+1) \pi,
~~~\hn= \left(\begin{array}{c}  0 \\
0 \\ -1 \end{array} \right).
\label{svortex}
\end{gather}
Obviously this becomes solutions of (\ref{bskyeq}). 
To see the physical content of the solutions, we can 
express $\hn$ in terms of the $CP^1$ field $\xi$ 
and find that the magnetic potential $C_\mu$ of 
the solution is given by
\begin{gather}
C_\mu=-2i\xi^\dagger\pd_\mu\xi
=2m \pd_\mu\varphi,  \nn\\
\xi= \left(\begin{array}{c}  0  \\
e^{im\varphi}  \end{array} \right), 
~~~\hn=\xi^\dagger\vec \sigma\xi
=\left(\begin{array}{c}  0 \\
0 \\ -1 \end{array} \right).
\label{hu1}
\end{gather}
Clearly this potential generates vanishing magnetic field  
everywhere, except the origin. But at the origin
the magnetic field becomes singular, and generates
the flux
\begin{gather}
\Phi=\oint_{\varrho\rightarrow \infty} C_\mu dx^{\mu}
=4\pi m.
\label{2mf}
\end{gather}
So they describe the singular magnetic vortex which 
carry the flux $4\pi m$ along the z-axis. The singular 
and regularized baby skyrmion solutions are shown 
in Fig. \ref{cbsky2}. 

One might have difficulty to understand these singular 
solutions, because (\ref{svortex}) tells that all field
configurations $\om$ and $\hn$ are trivial (i.e., 
constants). But as we have emphasized, the Skyrme 
theory has a hidden U(1) gauge symmetry. To reveal 
this symmetry we have to express $\hn$ by the $CP^1$ 
field $\xi$. And in terms of $\xi$, we  can show that 
(\ref{svortex}) has a non-trivial U(1) structure, as we 
have shown in (\ref{u1}). This is why (\ref{svortex}) 
describes the magnetic vortex. 

Of course, the quantization of the singular magnetic 
flux is topological. But the topology is not $\pi_2(S^2)$ 
of the baby skyrmion, but $\pi_1(S^1)$ of the hidden 
U(1). To see this notice that $\hn$ of (\ref{svortex}) 
defines the mapping $\pi_1(S^1)$ from the $S^1$ 
circle of constant radius in the $xy$-plane to 
the target space $S^1$ of the U(1) subgroup of 
SU(2) which leaves $\hn$ invariant \cite{ijmpa08}. 

The importance of this singular solution is that 
the popular baby skyrmions can be viewed as 
the regularized magnetic vortex, regularized by 
the non-trivial $f(\varrho)$ which satisfies the baby 
skyrmion equation (\ref{bskyeq}). So, just like 
the well known skyrmion is nothing but the regularized 
solution of the singular monopole (\ref{mono}) 
regularized by $\om$, the baby skyrmion is 
the regularized solution of the singular vortex 
(\ref{svortex}). 

Actually, there is another type of totally new magnetic 
vortex solutions. To see this notice
that
\begin{gather}
f(\varrho)= (2p+1) \frac{\pi}{2},
~~~\hn= \pm \left(\begin{array}{c}  \cos{m\varphi} \\
\sin{m \varphi} \\ 0 \end{array} \right),
\label{svortex1}
\end{gather}
also become singular solutions of (\ref{bskyeq}).
Clearly this describes the magnetic vortex given by
the potential $C_\mu$,
\begin{gather}
C_\mu=-2i\xi^\dagger\pd_\mu\xi
= m\pd_\mu\varphi,  \nn\\
\xi=\frac{1}{\sqrt 2} \left(\begin{array}{c}
1 \\ e^{im\varphi} \end{array} \right),
~~~\hn=\xi^\dagger\vec \sigma\xi,
\label{svortex1}
\end{gather}
which has the flux
\bea
\Phi=\oint_{\varrho\rightarrow \infty} C_\mu dx^{\mu}
=2\pi m.
\label{1mf}
\eea
This should be compared with (\ref{2mf}).

\begin{figure}
\begin{center}
\includegraphics[height=5cm, width=9cm]{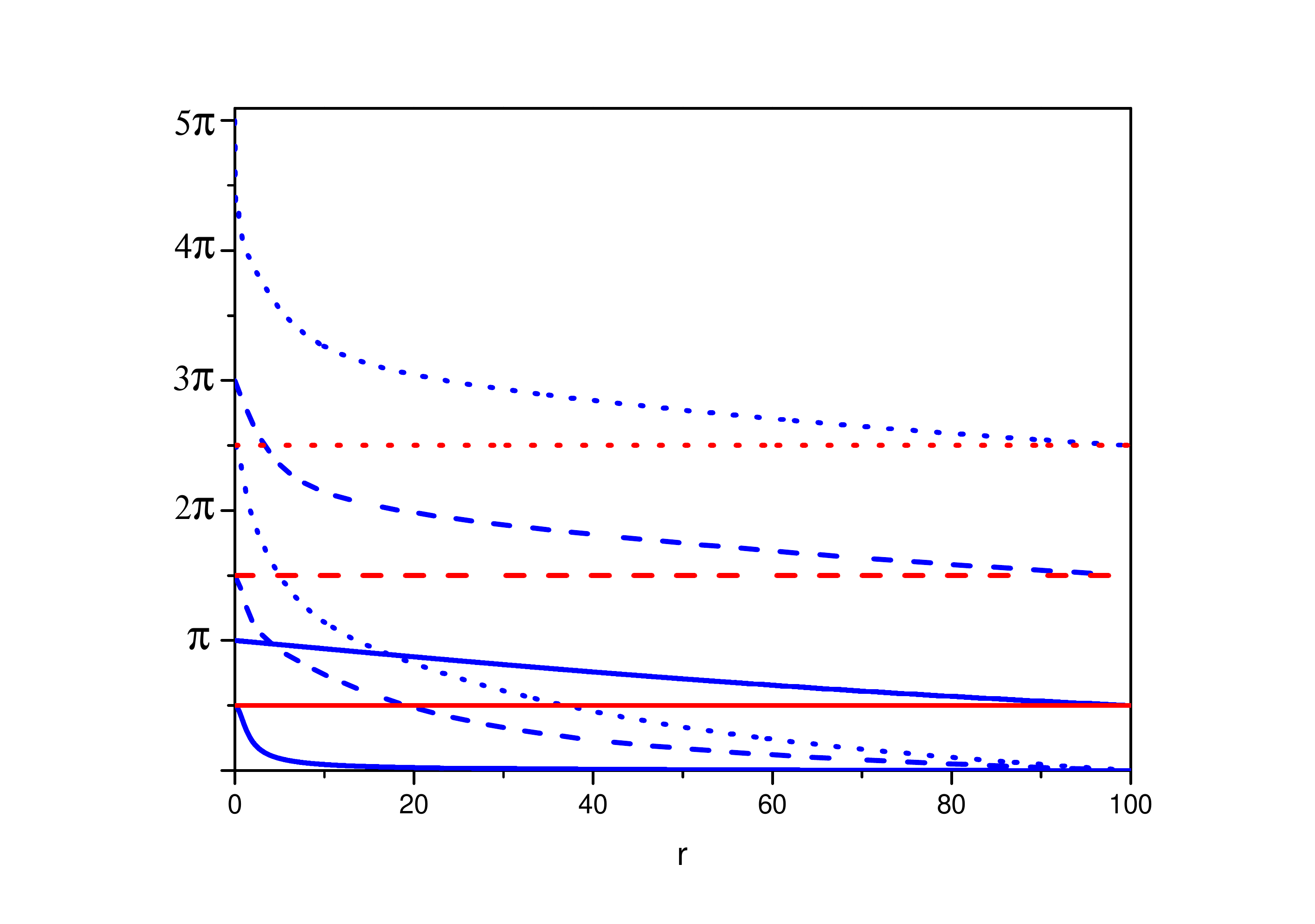}
\caption{\label{cbsky1} The singular and partly 
regularized baby skyrmion solutions which carry 
the $2\pi m$ magnetic flux for $m=1,2,3$. The red 
straight lines represent the singular solutions, 
and the blue curves represent the partly regularized 
solutions.}
\end{center}
\end{figure}

Moreover, they can also be regularized. To see this we
choose the ansatz (\ref{bskyans}) and the boundary
condition
\begin{gather}
f(0)=(2p+1) \frac{\pi}{2},~~~~f(\infty)=0.
\label{bskybc11}
\end{gather}
With this we can solve (\ref{bskyeq}) and obtain
the magnetic vortex solutions described by
the potential
\begin{gather}
C_\mu=-2i\xi^\dagger\pd_\mu\xi
=- m\cos f\pd_\mu\varphi, \nn\\
\xi= \left(\begin{array}{c}
\cos\frac{f(\varrho)}{2}e^{-\frac{i}{2}m\varphi} \\
\sin\frac{f(\varrho)}{2}e^{\frac{i}{2}m\varphi}
\end{array} \right),
~~~\hn=\xi^\dagger\vec \sigma\xi,
\label{rvortex11}
\end{gather}
which has the quantized magnetic flux along
the $z$-axis
\begin{gather}
\Phi =\Int H_{\varrho\varphi} d\varrho d\varphi
= 2\pi m.
\label{mqn1}
\end{gather}
Notice that, unlike (\ref{svortex1}), the magnetic 
potential of (\ref{rvortex11}) is regular at the origin. 

Furthermore, we have similar solutions with slightly
different boundary condition
\begin{gather}
f(0)=(2p+1) \pi, ~~~f(\infty)=(2p+1) \frac{\pi}{2}.
\label{bskybc12}
\end{gather}
and
\begin{gather}
C_\mu=-2i\xi^\dagger\pd_\mu\xi
=-m(\cos f +1)\pd_\mu\varphi, \nn\\
\xi= \left(\begin{array}{c}
\cos\frac{f(\varrho)}{2}e^{-im\varphi} \\
\sin\frac{f(\varrho)}{2}
\end{array} \right),
~~~\hn=\xi^\dagger\vec \sigma\xi.
\label{rvortex12}
\end{gather}
Again notice that the magnetic potential of these 
solutions is regular at the origin, and generates
$2\pi m$ flux. These solutions for $p=0,1,2$ are 
shown in Fig. \ref{cbsky1}.

The energy functional of these solutions is given by
\begin{gather}
E=2\pi\int \mathcal E \varrho d\varrho
=\pi\mu^2\int_0^{\infty}\Big[\Big(1
+\frac{\alpha^2}{\mu^2} \frac{n^2}{\varrho^2}
\sin^2f \Big)\dot f^2   \nn\\
+\frac{n^2}{\varrho^2}\sin^2f \Big]\varrho d\varrho.
\end{gather}
So the energy density of the solution (\ref{rvortex11})
is divergent at the orgin, but that of the solution
(\ref{rvortex12}) is divergent at the infinity. In this 
sense both of them are singular. 

\begin{figure}
\begin{center}
\includegraphics[height=5cm, width=9cm]{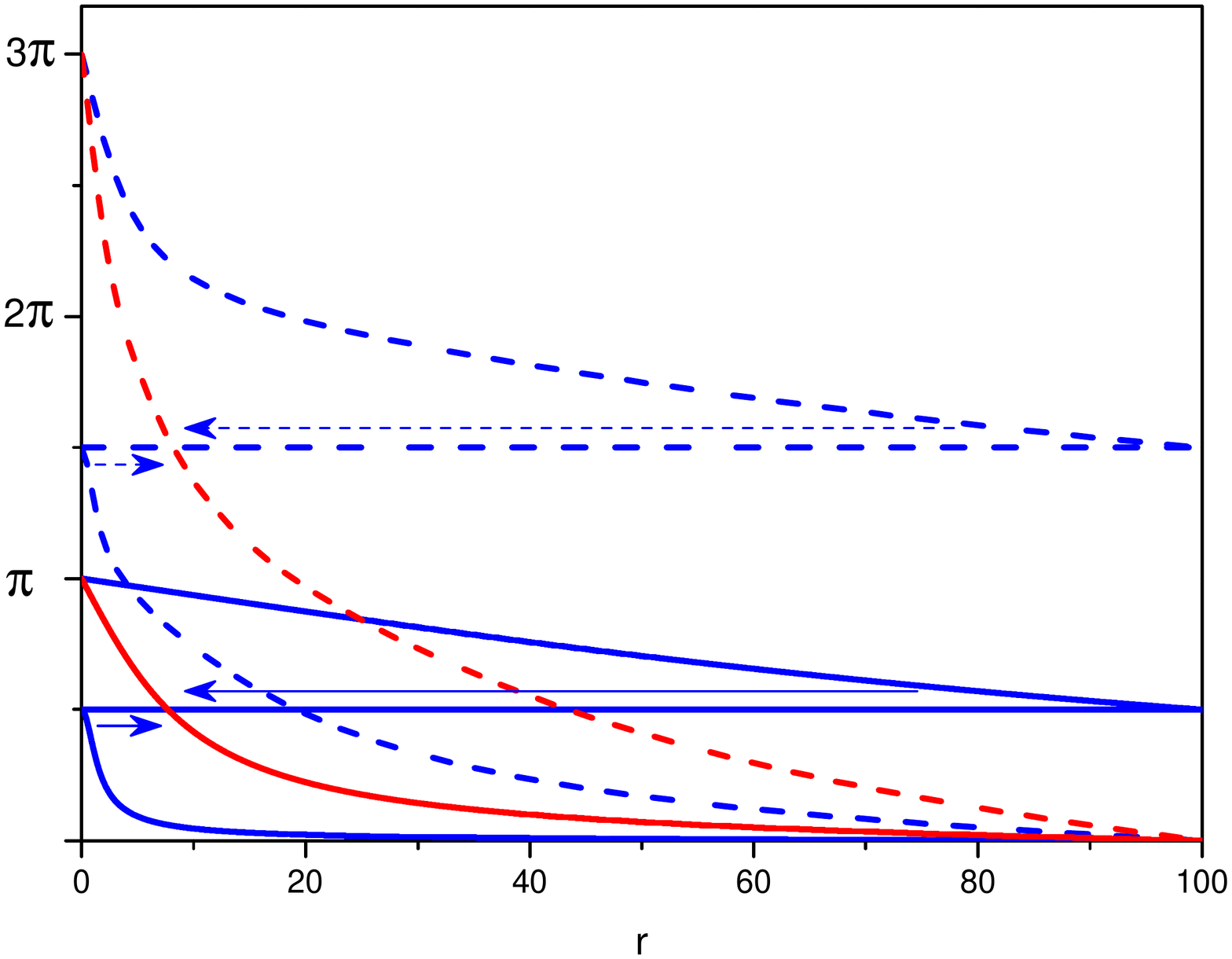}
\caption{\label{cbsky3} The regularized baby skyrmion
solution with $4\pi m$ flux made of two singular 
baby skyrmions with $2\pi m$ flux. Here we show 
only $m=1,2$ solutions for simplicity.}
\end{center}
\end{figure}

Notice, however, combining these two solutions with
the boundary condition (\ref{bskybc}) we obtain
the well known regular baby skyrmions. This is identical
to the observation that the skyrmions are the sum of
two half-skyrmions \cite{ijmpa08}. This is schematically 
shown in Fig. \ref{cbsky3}. The similarity between this 
figure and. Fig \ref{mosky} is unmistakable.

Obviously the new solutions have interesting features. 
But the important lesson that we learn from the above 
exercise for our purpose in this paper is that 
the solution (\ref{svortex1}) is the crux, the building 
block, of the baby skyrmions. All other solutions stem 
from this. This point becomes important for us to prove
the fact that the baby skyrmion can indeed be generalized 
to carry two topological numbers. 

To do that, notice first that the baby skyrmions
are described by $\hn$, without $\om$. And they carry 
only one topological number (the quantized magnetic 
flux) determined by $\hn$. So, by activating $\om$, 
we could obtain the generalized baby skyrmions which 
have two topological numbers. To show that this is 
possible, we activate $\om$ and generalize the ansatz 
(\ref{bskyans}) to
\begin{gather}
\om=\om(\varrho),
~~~\hn =\left(\begin{array}{c}
\sin{f(\varrho)}\cos{n\varphi} \\
\sin{f(\varrho)}\sin{n\varphi} \\
\cos{f(\varrho)}
\end{array} \right).
\label{cbskyans}
\end{gather}
With this the skyrme equation (\ref{skeq1}) is
reduced to
\begin{gather}
\Big(1+\frac{\alpha}{\kappa^2} \frac{n^2}{\varrho^2} \sin^2f\sin^2\frac{w}{2} \Big)\ddot f  \nn\\
+\Big[\frac{1}{\varrho}+\frac{\alpha}{\kappa^2}
\frac{n^2}{2\varrho^2} \dot f\sin 2f\sin^2\frac{w}{2}
-\frac{\alpha}{\kappa^2} \frac{n^2}{\varrho^3}
\sin^2f\sin^2\frac{w}{2}  \nn\\
+\Big(\cot\frac{w}{2}+\frac{\alpha}{\kappa^2}
\frac{n^2}{\varrho^2} \sin^2f\sin w \Big)
\dot w \Big]\dot f \nn\\
-\Big(\frac{n^2}{\varrho^2} 
+\frac{\alpha}{ \kappa^2} \frac{n^2}{4\varrho^2}
\dot w^2 \Big) \sin f\cos f=0,   \nn\\
\Big(1+\frac{\alpha}{\kappa^2} \frac{n^2}{\varrho^2} \sin^2f\sin^2\frac{w}{2} \Big)\ddot w \nn\\
+\Big(\frac{1}{\varrho}+
\frac{\alpha}{\kappa^2} \frac{n^2}{\varrho^2} \dot f \sin2f\sin^2\frac{w}{2}-\frac{\alpha}{\kappa^2}
\frac{n^2}{\varrho^3} \sin^2f\sin^2\frac{w}{2}   \nn\\
+\frac{\alpha}{ \kappa^2} \frac{n^2}{4\varrho^2}
\dot w \sin^2f\sin w \Big)\dot w   \nn\\
-\Big[\Big(1+ \frac{\alpha}{\kappa^2} \frac{2n^2}{\varrho^2} \sin^2f\sin^2\frac{w}{2} \Big)\dot f^2  \nn\\
- \frac{n^2}{\varrho^2} \sin^2f \Big] \sin w =0.
\label{genbskyeq}
\end{gather}
This is the generalized baby skyrmion equation in
Skyrme theory. Notice that, when $w(\varrho)=(2p+1)\pi$,
the first equation becomes exactly the baby skrymion
equation (\ref{bskyeq}).

\begin{figure}
\begin{center}
\includegraphics[height=5cm, width=9cm]{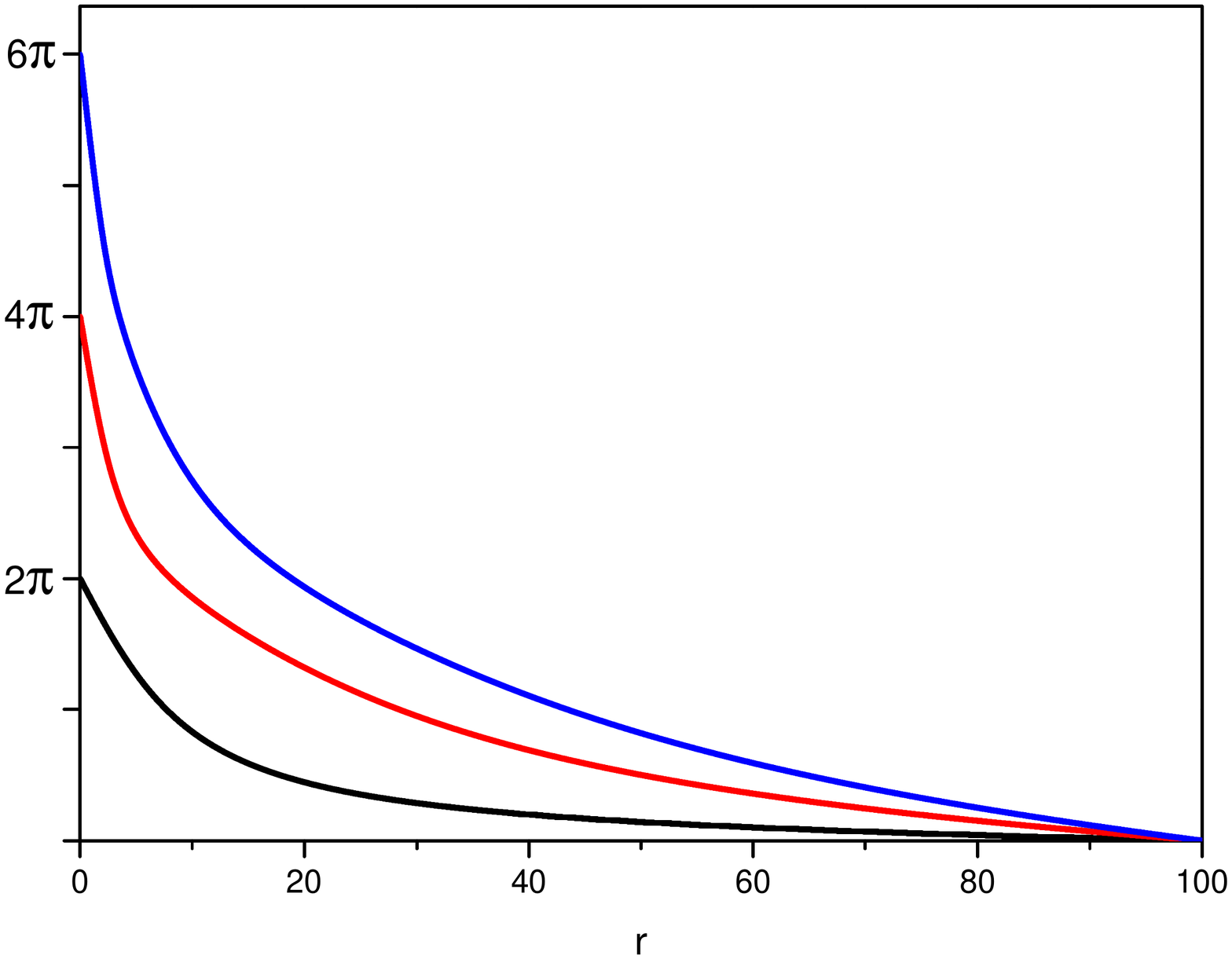}
\caption{\label{cbskymn} The generalized baby skyrmion
solutions which have two topological numbers $(m,n)$. 
Here $f$ is fixed ($f=(2p+1)\pi/2$), and we show 
the solutions for $\om$ for $n=1,2,3$.}
\end{center}
\end{figure}

This equation has a new set of vortex solutions. To see 
this, let $f(\varrho)=(2p+1) \pi/2$ as we did in 
(\ref{svortex1}). With this, the first equation of 
(\ref{genbskyeq}) is automatically satisfied, and
the second equation becomes
\begin{gather}
\Big(1+\frac{\alpha}{\kappa^2} \frac{n^2}{\varrho^2}
\sin^2\frac{w}{2} \Big)\ddot w \nn\\
+\frac{1}{\varrho}\Big(1+\frac{\alpha}{ \kappa^2}
\frac{n^2}{4\varrho } \dot w  \sin w
-\frac{\alpha}{\kappa^2}\frac{n^2}{\varrho^2}
\sin^2\frac{w}{2} \Big)\dot w \nn\\
- \frac{n^2}{\varrho^2} \sin w = 0.
\label{cbskyeq}
\end{gather}
Remarkably, this equation becomes exactly the baby
skyrmion equation (\ref{bskyeq}), if we identify $w(\varrho)=2f(\varrho)$. So, solving this with 
the boundary condition
\begin{gather}
\om(0)=2\pi n,~~~~~\om(\infty)=0, 
\label{cbskybc}
\end{gather}
we obtain the new generalized skyrmion solutions 
shown in Fig. \ref{cbskymn}.

Clearly they have singular $2\pi m$ magnetic 
flux, and have the energy 
\begin{gather}
E=\frac{\pi}{4}\mu^2\int_0^{\infty}\Big[\Big(1
+\frac{\alpha^2}{\mu^2} \frac{n^2}{\varrho^2} 
\sin^2\frac{w}{2} \Big)\dot w^2  \nn\\
+4\frac{n^2}{\varrho^2}\sin^2\frac{w}{2}
\Big]\varrho d\varrho.
\end{gather}
This tells that, unlike the singular baby skymion solution 
shown in (\ref{svortex1}), this solution becomes regular 
everywhere. 

Notice that asymptotically the baby skyrmion equation 
(\ref{bskyeq}) can be approximated to
\begin{gather}
\ddot f +\frac{1}{\varrho} \dot f  
- \frac{m^2}{\varrho^2} f \simeq 0,
\label{bskyeq1}
\end{gather}
so that the solutions shown in Figs. \ref{cbsky2}, 
\ref{cbsky1}, \ref{cbsky3} (and Fig. \ref{cbskymn}) 
have long $1/\varrho^m$ (or $1/\varrho^n$) tails, 
with no exponential damping. But this is a generic 
feature of the baby skyrmion \cite{ijmpa08,piet}. 

These solutions are interesting because they are new. 
But what is really important about these baby skyrmion 
solutions for our purpose here is that they have two 
independent topology, the flux topology $\pi_1(S^1)$ 
fixed by the invariant subgroup of $\hn$ and the shell 
(radial) topology $\pi_1(S^1)$ fixed by $\om$. 

This means that the baby skyrmions can also be 
generalized to carry two topological numbers,  
the monopole number $m$ of the magnetic flux 
fixed by $\hn$ and the winding number $n$ of 
the radial excitation $n$ fixed by $\om$. This 
confirms that the baby skyrmions (as well as 
the skyrmions) have two topology, and thus should 
be classified by two topological numbers.

\section{Multiple Vacua of Skyrme Theory}

Skyrme theory has been known to have rich topological
structures. It has the Wu-Yang type monopoles which have
the $\pi_2(S^2)$ topology, the skyrmions which have
the $\pi_3(S^3)$ topology, the baby skyrmions which have
the $\pi_1(S^1)$ topology, and the Faddeev-Niemi knots which
have the $\pi_3(S^2)$ topology \cite{prl01,plb04,ijmpa08}.
In the above we have shown that the theory has more
topological structure, and proved that the skyrmions
can be generalized to have two topological quantum
numbers.

Now we show that the Skyrme theory in fact has another
very important topological structure, the topologically
different multiple vacua. To see this, notice that
(\ref{skeq1}) has the solution
\bea
\om=2\pi p,~~~(p;~integer),
\label{skyvac}
\eea
independent of $\hn$. And obviously this is the vacuum
solution.

This tells that the Skyrme theory has multiple vacua
classified by the integer $p$ which is similar to
the Sine-Gordon theory. But unlike the Sine-Gordon
theory, here we have the multiple vacua without any
potential. Moreover, the above discussion tells that
the spherically symmetric skyrmions connect and occupy
the $p+1$ adjacent vacua. This means that we can
connect all vacua with the spherically symmetric skyrmions.
Of course, one could introduce such vacua in Skyrme theory
introducing a potential term in the Lagrangian \cite{nitta}.
This is not what we are doing here. We have these vacua
without any potential.

But this is not the end of the story. To see this notice that
(\ref{skyvac}) becomes the vacuum independent of $\hn$.
This means that $\hn$ can add the $\pi_3(S^2)$ topology
to each of the multiple vacua classified by another integer
$q$, because it is completely arbitrary. And this is precisely
the knot topology of the QCD vacuum \cite{plb07}.

Actually this is not be surprising. As we have already
emphasized, there is a deep connection between Skyrme
theory and QCD. So it is natural that the Skyrme theory
and QCD have similar vacuum structure. To clarify this
point we introduce a right-handed unit isotriplet
$(\hn_1,\hn_2,\hn_3=\hn)$, and impose the vacuum
isometry to the SU(2) gauge potential,
\bea
^{\forall_i} D_\mu \hn_i=0,
\eea
which assures $\vec F_\mn=0$. From this we obtain
the most general SU(2) QCD vacuum potential \cite{plb07}
\bea
&\A_\mu\rightarrow \hat \Omega_\mu
=-\dfrac12 \epsilon_{ijk} (\hn_i\cdot \pd_\mu \hn_j)
~\hn_k  \nn\\
&=\dfrac12 \epsilon_{ijk} (\hn_i\cdot \pd_\mu \hn_j)
~\hat e_k,
\label{qcdvac}
\eea
where $\hat e_1=(1,0,0),\hat e_2=(0,1,0),\hat e_3=(0,0,1)$.

This is the QCD vacuum which has the knot topology
$\pi_3(S^3)\simeq \pi_3(S^2)$. It is clear that
(\ref{qcdvac}) describes the $\pi_3(S^3)$ topology,
since $(\hn_1,\hn_2,\hn)$ defines the mapping
$\pi_3(S^3)$ from the compactified 3-dimensional
space to the SU(2) group space. But notice that
$(\hn_1,\hn_2,\hn)$ is completely determined by
$\hn$, up to the U(1) rotation which leaves $\hn$
invariant, which describes the mapping from the real
space $S^3$ to the coset space $S^2$ of $SU(2)/U(1)$.
So the QCD vacuum (\ref{qcdvac}) can also be classified
by the knot topology $\pi_3(S^2)$ \cite{plb07}.

Now it must be clear why the Skyrme theory has the same
knot topology. As we have noticed, the Skyrme theory
has the vacuum (\ref{skyvac}), independent of $\hn$.
But this $\hn$ (just as in the SU(2) QCD) defines
the mapping $\pi_3(S^2)$, with the $S^3$ compactification
of $R^3$, and thus can be classified by the knot topology.
Of course, in the Skyrme theory we do not need the
vacuum potential (\ref{qcdvac}) to describe the vacuum.
We only need $\hn$ which describes the knot topology.

This tells that the vacuum in Skyrme theory has the topology
of the Sine-Gordon theory and QCD combined together. This
means that the vacuum of the Skyrme theory can also be
classified by two quantum numbers $(p,q)$, the $\pi_1(S^1)$
of $\om$ and $\pi_3(S^2)$ of $\hn$. And this is so without
any extra potential. As far as we know, there is no other
theory which has this type of vacuum topology.

At this point we emphasize the followings. First, the knot
topology of $\hn$ is different from the monopole topology
of $\hn$. The monopole topology $\pi_2(S^2)$ is associated
to the isolated singularities of $\hn$, but the knot topology
$\pi_3(S^2)$ does not require any singularity for $\hn$.
And for a classical vacuum $\hn$ must be completely
regular everywhere. So only the knot topology, not the
monopole topology, can not describe a classical vacuum.
And this is precisely the vacuum topology of QCD.

Second, the knot topology of the vacuum is different
from the Faddeev-Niemi knot that we have in the Skyrme
theory \cite{prl01}. The Faddeev-Niemi knot is a unique
and real (i.e., physical) knot which carries energy, which
is given by the solution of (\ref{sfeq}). In particular,
we have the knot solution when $\om=(2n+1)\pi$. On the other
hand, we have the knot of the vacuum when $\om= 2\pi p$.
Moreover, the vacuum knot has no energy, and is not unique.
While the Faddeev-Niemi knot is unique, there are infinitely
many $\hn$ which describes the same vacuum knot topology.
So obviously they are different.

What is really remarkable is that the same $\hn$ has
multiple roles. It describes the monopole topology,
the knot topology of Faddeev-Niemi knot, and the knot
topology of the vacuum.

The fact that the Skyrme theory has topologically distinct
vacua raises more questions. Do we have the vacuum tunneling
in Skyrme theory? If so, what instanton do we have in this
theory? How different is it from the QCD instanton?

\section{Discussions}

The Skyrme theory was proposed as a low energy effective
theory of strong interaction where the baryons appear as
the topological solitons made of pions, the Nambu-Goldstone
field of the chiral symmetry breaking \cite{sky,witt,rho}.
This view has become the standard view \cite{man,bat,houg}.

But the Skyrme theory has many faces. As we have pointed
out, the theory can actually be viewed as a theory of monopole
which has the built-in Meissner effect \cite{prl01,plb04,ijmpa08}.
In fact all classical objects in Skyrme theory stem from
the monopole. Of course, the fact that the skyrmion is closely
related to the monopole has been well known. The rational
map $\pi_2(S^2)$ of $\hn$ which determines the baryon
number in the popular non spherically symmetric skyrmions
has been associated  to the monopole topology from
the beginning \cite{man,bat,houg}. But so far the fact
that the skyrmion itself becomes the monopole has not
been widely appreciated.

In this paper we have shown that the skyrmions are not
just the dressed monopoles but actually carry the monopole
number, so that they can be classified by two topological
numbers, the baryon number and the monopole number.
Moreover, we have shown that here the baryon number could
be replaced by the radial (shell) number, so that the skyrmions
can be classified by two topological numbers $(m,n)$,
the monopole number $m$ which describes the $\pi_2(S^2)$
topology of the $\hn$ field and the radial (shell) number
$n$ which describes the $\pi_1(S^1)$ topology of the $\om$
field. In this scheme the baryon number $B$ is given by
the product of two integers $B=mn$. This comes from
the following facts. First, the SU(2) space $S^3$ admits
the Hopf fibering $S^3\simeq S^2\times S^1$. Second,
the Skyrme theory has two variables, the angular variable
$\om$ which can represent the $\pi_1(S^1)$ topology and
the coset variable $\hn$ which represents the $\pi_2(S^2)$
topology.

In this view the popular (non spherically symmetric)
skyrmions are classified as the $(m,1)$ skyrmions, and
the radially excited spherically symmetric skyrmions
are classified as the $(1,n)$ skyrmions. and we can
construct the $(m,n)$ skyrmions adding the shell structure
to the $(m,1)$ skyrmions. Moreover, we have shown that
the skyrmions, when they are generalized to have two
topological numbers, should have the topological stability
of the two topology independently. 

Furthermore, we have shown that the baby skyrmions 
can also be generalized to carry two topology. 
Reactivating the radial variable $\om$, we have 
shown that the baby skyrmions in Skyrme-Faddeev 
theory can be generalized to new baby skyrmions in 
Skyrme theory. And this generalization naturally 
introduces a new $\pi_1(S^1)$ topology to the generalized 
solutions. This means that the new generalized baby 
skyrmions carry two topology, the shell (radial) 
topology$\pi_1(S^1)$ of $\om$ and the monopole topology 
$\pi_2(S^2)$ of $\hn$, and thus are classified by 
two topological numbers. This is remarkable.

As importantly, we have shown that the Skyrme theory
has multiple vacua. The vacuum of the theory has 
the structure of the vacuum of the Sine-Gordon theory 
and at the same time the structure of QCD vacuum. 
So the vacuum  can also be classified by two topological 
numbers $p$ and $q$ which represent the $\pi_1(S^1)$ 
topology of the $\om$ field and the $\pi_3(S^2)$ topology 
of the $\hn$ field.

The fact that the vacuum of the Skyrme theory has
the $\pi_1(S^1)$ topology is not surprising, considering
that it has the angular variable $\om$. Moreover,
the fact that the vacuum of the Skyrme theory has
the $\pi_3(S^2)$ topology of the QCD vacuum could easily
be understood once we understand that the Skyrme theory
is closely related to QCD. What is really remarkable
is that it has both $\pi_1(S^1)$ and $\pi_3(S^2)$ topology
at the same time. As far as we understand there is no
other theory which has this feature. This again is closely
related to the fact that $S^3$ admits the Hofp fibering
and that the theory has two variables $\om$ and $\hn$.

As we have remarked, our results raises more questions. Can
we extend the Faddeev-Niemi knots activating $\om$ to have
the shell structure? How can we obtain such solutions?
How about the vacuum tunneling in Skyrme theory? What
instanton do we have in this theory? The questions continue.

Clearly the above observations put the Skyrme theory in
a totally new perspective. Our results in this paper show
that the theory has so many new aspects which make
the theory more interesting. But most importantly our
results put the popular interpretation of the Skyrme theory
in an awkward position, and strongly imply that we need a
new interpretation of the Skyrme theory.

First of all, our result tells that the skyrmions carry two
topological numbers which are conserved independently.
So, if the skyrmions become the baryons, the baryons
must have two conserved quantities. This is very difficult
to accommodate, because the baryons seem to have no
conserved quantity other than the baryon number.

Moreover, the Skyrme theory have the (generalized) baby 
skyrmions which can be viewed as skyrmion-antiskyrmion 
pair infinitely separated apart. So, if the skyrmions become 
the baryons, we must have the string-like hadrons. 
In fact, the theory has too many topological solitons
which could be interpreted as strongly interacting 
particles. It has not only the skyrmions but also 
the Faddeev-Niemi knots. In this case the knots should 
be interpreted as mesons made of baryon-antibaryon 
pair. 

But the knots are expected to have huge mass, roughly 
about 10 times heavier than the proton mass, simply 
because of the geometricshape \cite{plb04,ijmpa08}. 
So, if the skyrmion is treated as the baryon, we must 
accept the existence of a topologically stable meson 10 
times heavier than proton. And we should take this 
possibility seriously.

All of these evidences strongly suggest that the Skyrme
theory need a totally new interpretation. It simply has
too many interesting featutres to be considered as 
a theory of strong interaction. These new features and 
the questions raised in this paper will be discussed in 
a separate publication \cite{cho}.

{\bf ACKNOWLEDGEMENT}

~~~The work is supported in part by the National 
Natural Science Foundation of China (Grant 11575254), 
Chinese Academy of Sciences Visiting Professorship 
for Senior International Scientists (Grant 2013T2J0010), 
Chinese Scholarship Council, National Research 
Foundation of Korea funded by the Ministry of Education 
(Grants 2015-R1D1A1A0-1057578 and 
2015-R1D1A1A0-1059407), and by Konkuk University.

\end{document}